\documentclass[10pt,journal,compsoc]{IEEEtran}
\usepackage[T1]{fontenc}

\usepackage{cite}
\usepackage[colorlinks,linkcolor=black,citecolor=black, urlcolor=black]{hyperref}
\usepackage{soul}

\usepackage{graphicx}

\usepackage{amssymb}
\usepackage{amsmath}
\usepackage{bm}

\usepackage[linesnumbered,ruled]{algorithm2e}
\usepackage{booktabs}
\usepackage{subfigure}

\begin{document}

\title{Enabling Feedback-Free MIMO Transmission for FD-RAN: A Data-driven Approach}

\author{Jingbo~Liu,~\IEEEmembership{Student~Member,~IEEE,}
Jiacheng~Chen,~\IEEEmembership{Member,~IEEE,}
Zongxi~Liu,~\IEEEmembership{Student~Member,~IEEE,}
and~Haibo~Zhou,~\IEEEmembership{Senior~Member,~IEEE}


\IEEEcompsocitemizethanks{
\IEEEcompsocthanksitem J. Liu is with the School of Electronic Information and Electrical Engineering, Shanghai Jiao Tong Universtiy, Shanghai 200240, China, and also with the Peng Cheng Laboratory, Shenzhen 518000, China.
\protect\\
E-mail: liujingbo@sjtu.edu.cn.

\IEEEcompsocthanksitem J. Chen (Corresponding author) is with the Department of Strategic and Advanced Interdisciplinary Research, Peng Cheng Laboratory, Shenzhen 518000, China.
\protect\\
E-mail: chenjch02@pcl.ac.cn.

\IEEEcompsocthanksitem Z. Liu and H. Zhou are with the School of Electronic Science and Engineering, Nanjing University, Nanjing 210023, China.
\protect\\
E-mail: zongxiliu@smail.nju.edu.cn, haibozhou@nju.edu.cn.

} 

}


\IEEEtitleabstractindextext{
\begin{abstract}

To enhance flexibility and facilitate resource cooperation, a novel fully-decoupled radio access network (FD-RAN) architecture is proposed for 6G.
However, the decoupling of uplink (UL) and downlink (DL) in FD-RAN makes the existing feedback mechanism ineffective.
To this end, we propose an end-to-end data-driven MIMO solution without the conventional channel feedback procedure.
Data-driven MIMO can alleviate the drawbacks of feedback including overheads and delay, and can provide customized precoding design for different BSs based on their historical channel data.
It essentially learns a mapping from geolocation to MIMO transmission parameters.
We first present a codebook-based approach, which selects transmission parameters from the statistics of discrete channel state information (CSI) values and utilizes integer interpolation for spatial inference.
We further present a non-codebook-based approach, which 1) derives the optimal precoder from the singular value decomposition (SVD) of the channel; 2) utilizes variational autoencoder (VAE) to select the representative precoder from the latent Gaussian representations; and 3) exploits Gaussian process regression (GPR) to predict unknown precoders in the space domain.
Extensive simulations are performed on a link-level 5G simulator using realistic ray-tracing channel data.
The results demonstrate the effectiveness of data-driven MIMO, showcasing its potential for application in FD-RAN and 6G.

\end{abstract}

}

\maketitle

\section{Introduction} \label{intro}
\IEEEPARstart{C}{ommercialization} of the fifth generation (5G) mobile network is ongoing worldwide, yet some fundamental challenges have not been fully addressed.
Firstly, the scarce spectrum resources are still not efficiently utilized.
Secondly, it is still difficult to guarantee users' quality of experience (QoE).
Thirdly, the costs of network infrastructure and operation are very high.
To address the above issues, the fully-decoupled radio access network (FD-RAN) architecture is proposed for the upcoming sixth generation (6G) mobile network \cite{wang2023road}.
It decouples the base stations (BSs) into control BSs (C-BSs) and data BSs, while data BSs are further decoupled into downlink (DL) BSs and uplink (UL) BSs\cite{yu2019fully, yu2022fully, zhao2023fully}.
With control, UL, and DL fully separated, control and resource allocation flexibility can be exploited to meet the requirements of 6G \cite{qian2023enabling, liu2023leveraging, xu2023federated}.

However, such a decoupled architecture makes the existing feedback schemes employed in 5G ineffective.
In 5G, take DL transmission as an example, BS first transmits channel state information reference signals (CSI-RSs) to user equipment (UE).
Then, UE estimates the channel and calculates the transmission parameters, which are fed back to the BS for carrying out multiple-input and multiple-output (MIMO) transmission.
In FD-RAN, DL BS is unable to obtain the CSI feedback from UE directly, since UL and DL BSs are physically separated.
It is also not efficient to use the control BS for CSI feedback, as the resource consumption for all UEs is high, and the additional delay incurred by control BS will make the feedback inaccurate.
Thus, how to implement MIMO transmission without CSI feedback is a significant challenge for FD-RAN.

On the other hand, the state-of-the-art 5G MIMO techniques also suffer from issues caused by feedback.
5G adopts closed-loop spatial multiplexing (CLSM) as its main MIMO scheme \cite{36.213}, due to its high capacity than open-loop spatial multiplexing (OLSM) and transmit diversity, as well as its ability to exploit massive antennas.
Specifically, the feedback, i.e., CSI, includes precoder, number of spatial streams, and modulation and coding scheme (MCS), which are necessary transmission parameters for MIMO.
When the number of transmit antenna increases, the overheads of pilots, namely CSI-RSs, and CSI feedback become much higher, reducing the actual data rate.
Also, feedback delay will affect the transmission performance, especially when the channel propagation environment changes fast.

To this end, we propose a totally different solution, namely data-driven MIMO.
Instead of relying on feedback to acquire necessary channel information for MIMO, we exploit the knowledge from historical channel data and only use UE's geolocation to determine transmission parameters.
Intuitively, data-driven MIMO is feasible because the channel and the corresponding transmission parameters are strongly correlated with geolocation since the signal propagation environment at a certain location will not dramatically change.
Although the channel is generally time-varying, we expect performance loss can be mitigated by saving overheads of pilots and CSI feedback.
Data-driven MIMO falls in the broader field of zero-feedback transmission, which has been investigated in literature \cite{bletsas2010simple, alexandris2014reachback, suleiman2017review}.

In this paper, we attempt to realize the same capability as CLSM, namely MIMO spatial multiplexing.
We adopt an end-to-end approach, so we need to construct a mapping from location to transmission parameters.
Since we do not have any channel feedback, the mapping is fixed, or we can say it is static in the time domain.
Given that the mapping can only be trained with existing channel data, we need to find the transmission parameters of all the possible locations.
Then, inference in the space domain is required.
It is also similar for the frequency domain if multiple subcarriers are considered.

Data-driven MIMO is very challenging since the performance loss incurred by fixed transmission parameters is definite and should be reduced as much as possible, to make the approach meaningful.
One major challenge is to deal with the channel characteristics in time, frequency, and space domains.
Specifically, in time domain we need to find the "representative" transmission parameters considering the channel fluctuations, and in space and frequency domains we need to infer the transmission parameters considering the channel correlations.
Another key challenge is how to jointly choose transmission parameters, namely precoder, rank indicator (RI), and channel quality indicator (CQI).
Generally, the precoder and RI are determined at the same time, since RI indicates the rank of the precoder.
Then, proper CQI can be derived based on the chosen precoder and RI.
Any improper choice of these three parameters would lead to performance degradation, especially considering RI and CQI are quantized discrete values.
On the one hand, low RI and CQI values will lead to loss in transmission performance.
On the other hand, high RI values will bring more interference among data streams while high CQI values can lead to increase in bit error ratio (BER), both of which will lower the performance.
Existing works mainly concentrate on beamforming and use only one stream.
\cite{sklivanitis2013testbed} studies the beamforming problem without feedback, and \cite{nguyen2022deep} proposes a deep learning framework for beam selection.
Different from them, we jointly consider the three transmission parameters, as in 5G CLSM.

To address the above challenges, we propose two approaches to realize data-driven MIMO for FD-RAN DL transmission.
The first approach is based on 5G Type \uppercase\expandafter{\romannumeral1} codebook \cite{38.214}, which basically quantizes the precoder and assigns each precoder with a precoding matrix indicator (PMI).
Thus, all the transmission parameters are discrete values and we can use a statistic-based method to choose the fixed transmission parameters and further utilize integer interpolation for spatial prediction.
However, the quantization of the codebook also leads to the inaccuracy of the precoder.
Based on the fact that singular value decomposition (SVD) of the channel matrix can optimally calculate the precoder, we further consider a second approach that starts processing from the optimal precoders.
To fix the transmission parameters in the time domain, we exploit variational autoencoder (VAE) \cite{kingma2013auto} to discover the latent representations of optimal precoders in lower dimensions.
The encoder of VAE returns a distribution over the latent space, and close sample points are expected to have similar representations in the latent space, which makes the results of the generative process of decoder more meaningful.
Given the latent representations are Gaussian variables, we further utilize Gaussian progress regression (GPR) for spatial prediction.
Note that in this paper, we consider transmission on narrow bands, and do not infer in the frequency domain, thus the same transmission parameters are applied to all subcarriers.
To summarize, the contributions of this paper are listed as follows:
\begin{itemize}
    \item
    We propose a novel channel feedback free solution to realize the advanced MIMO spatial multiplexing for both FD-RAN and future 6G.
    The proposed data-driven MIMO learns from the historical channel data and only utilizes geolocation to determine all the transmission parameters in an end-to-end way.
    The data-driven approach essentially customizes the MIMO transmission at each BS.
    \item
    We propose an easy-to-implement codebook-based approach for data-driven MIMO.
    Transmission parameters are selected using statistics of CSI values, and integer interpolation is used to handle the correlation of channel in the space domain.
    \item
    We further propose an SVD-based approach for performance improvement.
    VAE is used to select the representative precoder in time domain from latent Gaussian representations.
    Then, GPR is used to predict the unknown precoders in space domain.
    Natural neighbor interpolation (NNI) is utilized to infer CQIs at unknown locations.
    \item
    We perform extensive simulations based on a 5G-compatible link-level simulator and realistic ray-tracing channel data.
    Simulation results demonstrate the effectiveness of the proposed two
    approaches, and also the superiority of the SVD-based approach in terms of spatial inference.
    Compared with state-of-the-art 5G CLSM, we demonstrate the potential of data-driven MIMO, considering the saved overheads of pilots and feedback.

\end{itemize}

The remainder of this paper is organized as follows.
Related works are introduced in Section \ref{RW}.
The system model and problem formulation are given in Section \ref{SM and PF}.
The codebook-based approach and SVD-based approach are presented in Section \ref{CA} and Section \ref{SA}, respectively.
Simulation results and discussions are given in Section \ref{SR and D}.
Finally, Section \ref{Con} concludes the whole paper.

\section{Related Works}
\label{RW}

Due to heavy overheads of feedback in massive MIMO, a lot of efforts have been made to reduce the overheads to unleash the potential of 5G
\cite{guo2022overview, wen2018deep, wang2018deep, guo2020convolutional, chen2021deep}.
A comprehensive review of deep learning-based CSI feedback schemes is presented in \cite{guo2022overview}.
In \cite{wen2018deep}, a CSI compressing and recovering mechanism called CsiNet is first proposed for reducing the CSI feedback.
It treats the channel in angle and delay domains as an image, and exploits an encoder to compress the channel at UE side and a decoder to recover it at BS side.
To deal with the time-varying characteristics of channel, long short-term memory (LSTM) is employed for CsiNet \cite{wang2018deep}, which can achieve a better tradeoff between complexity and compression ratio.
Then, convolutional neural networks are further used for compressing CSI \cite{guo2020convolutional}, with quantization of it taken into consideration.
To cope with the existing feedback mechanism based on codebook, \cite{chen2021deep} proposes a bidirectional LSTM-based CSI feedback architecture to achieve a fine mapping between the precoder and corresponding PMI, where Type \uppercase\expandafter{\romannumeral1} and Type \uppercase\expandafter{\romannumeral2} codebooks of 5G are taken into account.
Different from the works mentioned above, we study the data-driven MIMO that requires no channel feedback at all.

Furthermore, a number of studies have been made to eliminate the channel feedback \cite{vasisht2016eliminating, chikha2022radio, alrabeiah2019deep, mismar2019deep, hanna2022destination}.
In \cite{vasisht2016eliminating}, a system called R2-F2 that enables BS to predict the downlink channel based on the estimated uplink channel is introduced.
R2-F2 essentially builds up a channel-to-path transform that captures the underlying physical paths to infer the channels at different frequency bands, which provides a 0.7 dB of beamforming gain.
The problem of channel mapping in space and frequency domains is studied in \cite{alrabeiah2019deep}.
Utilizing a fully connected network architecture, the channel at a certain location and frequency band can be mapped into the one at a different location and frequency band.
In \cite{chikha2022radio}, a radio environment map (REM) of one BS is built on the measurements from all grids of beams.
Then, the reference signal received powers utilizing different beamforming schemes can be derived based on REMs, facilitating the interference coordination between the serving and neighboring cells.
The problem of joint beamforming, power control, and interference coordination is considered in \cite{mismar2019deep}.
A deep reinforcement learning architecture is proposed at both sub-6 GHz and millimeter wave frequency bands, where UE merely sends back its received signal to interference plus noise ratio (SINR) and coordinates to BS.
In \cite{hanna2022destination}, the authors propose a destination-feedback-free distributed beamforming scheme, where location information is required.
One radio acts as a leader and guides the others to point toward the destination.
Different from the above works, we aim to determine all three transmission parameters for any set of channel matrices exploiting the historical channel data.

Owing to VAE's capability to capture the underlying characteristic of inputs and generate meaningful outputs, it has drawn the attention of numerous scholars and has been adopted for a wide range of areas \cite{han2019multi, lin2020video, liang2021variational, zhou2021vae, hussien2022prvnet, liu2021fire}.
Han \emph{et al.} utilize VAE to learn the features of 3D point clouds from multiple angles \cite{han2019multi}, where recurrent neural network (RNN) is used to learn each local geometry.
The proposed method can effectively catch the global and local features of point clouds based on the multi-angle analysis.
In \cite{lin2020video}, a modified VAE architecture is proposed for video segmentation and tracking.
It consists of one encoder and three decoders, and a GP model is introduced to strengthen the latent representation of VAE.
\cite{liang2021variational} applies VAE to data analytics in the area of Internet of Things (IoT), where transfer entropy is introduced to VAE's loss function to obtain a more regularized latent space and less reconstruction error.
In \cite{zhou2021vae}, a deep support vector data description (SVDD) based on VAE is proposed for anomaly detection, where SVDD and VAE are jointly optimized to derive the separated latent representations.
Meanwhile, VAE has also been tried to be applied in mobile communications.
In \cite{hussien2022prvnet}, VAE is employed to reduce the CSI feedback where an additive Gaussian white noise (AWGN) channel is assumed.
In \cite{liu2021fire}, FIRE, an end-to-end approach empowered by VAE, is proposed to obtain the downlink channel in frequency division duplex (FDD) system without any feedback from UE.
Instead of recovering the inputs like traditional VAE, FIRE learns the mapping between the uplink channel and the corresponding downlink channel, thereby enabling the acquisition of the downlink channel at BS.
Different from the works mentioned above, we exploit VAE to reduce the dimension of precoders to find the representative precoder in time domain and generate meaningful precoders for spatial inference.

\section{System Model and Problem Formulation}
\label{SM and PF}
In this section, we first introduce the system model of DL MIMO transmission in FD-RAN.
Then, we analyze and formulate the problem of data-driven MIMO.

\subsection{System Model}

\begin{figure}[t]
    \begin{center}
        \includegraphics[width=0.45\textwidth]{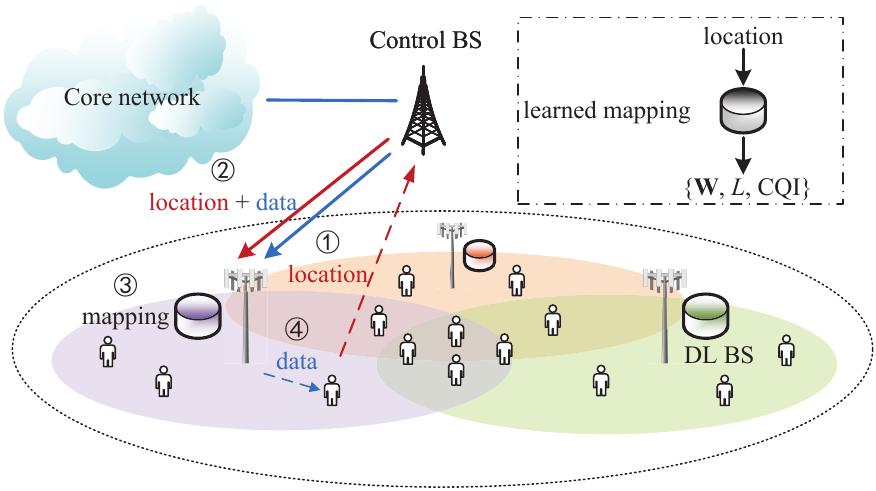}
        \caption{System model.}
        \label{scenario}
    \end{center}
\end{figure} \par

In this paper, we study the DL single-user MIMO (SU-MIMO) transmission of FD-RAN.
Fig. \ref{scenario} illustrates the system model, where a control BS coordinates with a single DL BS and multiple UEs.
The control BS obtains the locations reported by users and informs the DL BS.
Then, exploiting the mapping, DL BS can find out transmission parameters and carry out MIMO transmission based on UE's geolocation.
Although the feedback of location information through control BS also has delay, the influence on the transmission performance is limited.
On the one hand, the user's location will not significantly change in seconds in typical scenarios.
Also, the fixed transmission parameters will not change dramatically at close locations.
On the other hand, the user's location is usually easy to predict when it changes fast, e.g., when the user is inside a car or high-speed train.
Meanwhile, integrated sensing and communication (ISAC) \cite{liu2022integrated} can be utilized to sense the location of UE.
In this paper, we only focus on the performance at static user locations.

The whole signal processing procedure is based on orthogonal frequency division multiplexing (OFDM) and MIMO technologies.
OFDM is a multicarrier modulation mode employed in 4G long-term evolution (LTE) and is still a major modulation scheme in 5G new radio (NR).
Leveraging the discrete Fourier transform (DFT), OFDM transforms the frequency selective broadband channel into $K$ flat narrowband channels, and $k \in \mathcal{K}= \{1, \ldots, K\}$ refers to the $k$-th orthogonal subcarrier.
The DL BS and UE are assumed to be equipped with $N_{tx}$ transmit antennas and $N_{rx}$ receive antennas, respectively.
Then the received symbol vector $\bm{c}_{k} \in \mathbb{C}^{N_{rx} \times 1}$ on the subcarrier $k$ at any sampling time instant can be expressed as
\begin{equation}
    \bm{c}_{k} = \mathbf{H}_{k}\mathbf{W}\bm{s}_{k} + \bm{n}_{k}, \quad k \in \mathcal{K},
\end{equation}
where $\mathbf{H}_{k} \in \mathbb{C}^{N_{rx} \times N_{tx}}$ is the channel matrix that the transmitted symbol vector $\bm{s}_{k} \in \mathbb{C}^{L \times 1}$ with normalized unit power experiences on the subcarrier $k$.
$\bm{n}_{k} \sim \mathcal{CN}(\bm{0}, \varsigma^2_n\mathbf{I}_{N_{rx}})$ is the additive zero means complex-valued white Gaussian noise with variance $\varsigma^2_n$, where $\mathbf{I}_{N_{rx}}$ is the identity matrix of dimension $N_{rx} \times N_{rx}$.
$\mathbf{W} \in \mathbb{C}^{N_{tx} \times L} $ is the employed precoding matrix and $L \in \{ 1, \ldots, \max(L) = \Upsilon = \min\{N_{rx}, N_{tx}\} \} $ is the number of employed layers for spatial multiplexing.
Note that since we do not infer in the frequency domain, the same precoder is used for all subcarriers.
How to determine the appropriate precoder will be introduced in the following sections.
RI corresponds to the rank of precoder $\mathbf{W}$, which equals to $L$.

At the receiver side, $\mathbf{H}_{k}\mathbf{W}$ and noise variance $\bm{n}_{k}$ are estimated by demodulation reference signals (DM-RSs).
Then, the received symbol vector $\bm{c}_{k}$ is equalized by an equalizer, which is denoted by $\mathbf{E}_{k} \in \mathbb{C}^{L \times N_{rx}}$.
The post-equalization received symbol vector $\bm{m}_{k} \in \mathbb{C}^{L \times 1}$ is written as
\begin{equation}
    \label{post-equalization}
    \bm{m}_{k} = \mathbf{E}_{k}\bm{c}_{k} = \mathbf{E}_{k}\mathbf{H}_{k}\mathbf{W}\bm{s}_{k} + \mathbf{E}_{k}\bm{n}_{k}, \quad k \in \mathcal{K}.
\end{equation}
We denote the multiplication of three matrices $\mathbf{E}_{k}\mathbf{H}_{k}\mathbf{W}$ as $\mathbf{G}_{k} \in \mathbb{C}^{L \times L}$, which can be seen as the equivalent channel between transmitter and receiver.
Then SINR of $\ell$-th layer after equalization is given as
\begin{equation}
    \label{SINR}
    \text{SINR}_{k,\ell} = \frac{|\mathbf{G}_{k}(\ell,\ell)|_{F}}
    {\sum_{i=1, i \neq \ell}^{L} |\mathbf{G}_{k}(\ell,i)|_{F} + \varsigma^2_n \sum_{i=1}^{L} |\mathbf{E}_{k}(\ell,i)|_{F}},
\end{equation}
where $k \in \mathcal{K}, \ell \in \mathcal{L} = \{1,\ldots,L\}$ and $|\cdot|_{F}$ denotes the Frobenius norm.
$\mathbf{G}_{k}(\ell,i)$ denotes the element of the equivalent channel $\mathbf{G}_{k}$ that lies in the $\ell$-th row and $i$-th column.
Thus, the numerator of (\ref{SINR}) represents the desired signal power of layer $\ell$ on subcarrier $k$.
The first term of the denominator refers to the inter-layer interference and the second term is the enhanced noise.

Then, the post-equalization SINRs over all subcarriers $\mathcal{K}$ and all layers $\mathcal{L}$ are mapped into one effective SINR of an equivalent single input and single output (SISO) system with AWGN  channel\cite{brueninghaus2005link}, which has similar transmission performance with the MIMO OFDM system.
The mapping is defined as
\begin{equation}
    \label{effSINR}
    \text{SINR}_{\text{eff}} = \alpha f^{-1}\left(\frac{1}{KL} \sum_{k \in \mathcal{K}} \sum_{\ell \in \mathcal{L}} f\left(\frac{\text{SINR}_{k,\ell}}{\alpha}\right)\right),
\end{equation}
where $f$ denotes the mapping function and $f^{-1}$ is its inverse. Here, the mutual information effective SNR mapping (MIESM) is adopted, and $f$ is the bit interleaved coded modulation (BICM) capacity \cite{caire1996capacity}.
$\alpha$ is the adjustment factor that adapts the block error rate (BLER) performance of the equivalent channel to approximate that of the original channel as much as possible.
At last, the CQI value is chosen to be the maximal MCS value that keeps BLER lower than 0.1.
A high MCS value means that the high-order modulation and high coding rate scheme are employed, which brings high spectrum efficiency and throughput.

\subsection{Problem Formulation}

In essence, the problem of data-driven MIMO is to estimate a mapping from geolocation to transmission parameters, i.e., $ \bm{\varrho} = (x,y,z) \rightarrow \{\mathbf{W}, L, \text{CQI}\}$ based on the channel data.
$x$, $y$ and $z$ are Cartesian coordinates in Euclidean space.
The objective is to counter performance loss caused by fixing transmission parameters in the time domain and inferring unknown transmission parameters in the space domain.

We consider the coverage area of a DL BS and we assume that the environment will not change significantly over the studied long time span.
Specifically, the existing channel data is a set of channel matrices $\mathcal{H}$, where every element is a channel matrix $\mathbf{H}_{t,k,q}$ with dimension $N_{rx} \times N_{tx}$.
$t \in \mathcal{T} = \{ 1, \ldots, T \}$ refers to index in the time domain.
Note that $t$ does not need to be a real time point, since we will not do any inference in the time domain. Different values of $t$ only indicate that the channel matrices are different due to the time-varying characteristics of the channel.
$k$ denotes the $k$-th subcarrier, which relates to the channel in the frequency domain.
$q \in \mathcal{Q} = \{1, \ldots, Q\}$ is the index of location, which corresponds to the channel sample in the space domain.
Therefore, $|\mathcal{H}| = T \times K \times Q$.

In data-driven MIMO, utilizing the mapping derived from existing channel data in $\mathcal{H}$, we aim to maximize the throughput at any time and location sample in the coverage area of DL BS.
The problem can be formulated over an arbitrary set of channel matrices, e.g., $\mathcal{H}'$:
\begin{equation}
    \label{problem}
    \begin{split}
           &\max_{\{\mathbf{W}_{q}, L_{q}, \text{CQI}_{q}\}}
    \sum_{t \in \mathcal{T}'} \sum_{k \in \mathcal{K}} \sum_{q \in \mathcal{Q}'} \mathfrak{T}(\mathbf{H}_{t,k,q}, \{\mathbf{W}_{q}, L_{q}, \text{CQI}_{q}\}) \\
    & \text{s.t.} \left\{
    \begin{split}
        & \mathbf{H}_{t,k,q} \in \mathcal{H}', \quad
        \forall t \in \mathcal{T}', \forall k \in \mathcal{K}, \forall q \in \mathcal{Q}'\\
        & \|\mathbf{W}_{q}\|_{F} = 1, \quad
        \forall q \in \mathcal{Q}' \\
        & L_{q} = r(\mathbf{W}_{q}) \in\{1, \ldots, \Upsilon\},  \quad
        \forall q \in \mathcal{Q}'\\
        & \text{CQI}_{q} \in \{1,\ldots,15\}, \quad
        \forall q \in \mathcal{Q}',
    \end{split}
    \right.
    \end{split}
\end{equation}
where $\mathfrak{T}$ denotes the throughput, and it depends on the channel and transmission parameters.
Here, throughput is not calculated on the theoretical formula.
$\|\mathbf{W}_{q}\|_{F} = 1$ indicates the power of precoder is normalized to 1.
$r(\cdot)$ refers to the rank of a matrix.
CQI is an integer, ranging from 1 to 15 \cite{38.214}.
Since we fix the transmission parameters in time and frequency domains, we only use the subscript $q$ to denote the transmission parameters at location $q$, which are the same for any time point $t$ and subcarrier $k$.

\section{Codebook-based Approach}
\label{CA}
In this section, we first introduce Type \uppercase\expandafter{\romannumeral1} codebook used in 5G CLSM and then present the approach based on the statistic of historical CSI.

\subsection{Codebook-based 5G CLSM}
According to 5G NR standard \cite{38.214}, Type \uppercase\expandafter{\romannumeral1} codebook is employed for SU-MIMO and varies when BS employs single-panel or multi-panel antennas.
In this paper, we assume DL BS is equipped with a single-panel antenna array of configuration $(N_{1}, N_{2})$ with cross-polarization, where $N_{1}$ and $N_{2}$ represent the number of horizontal and vertical antenna ports in the direction of one polarization, respectively.
Thus, with two polarizations combined together, $N_{tx} = 2N_{1}N_{2}$.

Considering the tradeoff between the overheads of feedback and link performance, Type \uppercase\expandafter{\romannumeral1} codebook is designed in accordance with the selection of two-dimensional (2D) DFT beams.
In addition to the original $(N_{1}, N_{2})$ spatial beams, Type \uppercase\expandafter{\romannumeral1} codebook oversamples the horizontal and vertical beams by factors $O_{1}$ and $O_{2}$ to obtain a finer granularity.
Then, the spatial orthonormal bases of horizontal dimension $\bm{a}_{1}$ and vertical dimension $\bm{a}_{2}$ can be expressed as
\begin{equation}
    \label{spatial basis}
    \begin{split}
        \bm{a}_{1} &= \left[1, e^{j\frac{2\pi\theta_{1}}{N_{1}O_{1}}}, \ldots, e^{j\frac{2\pi\theta_{1} (N_{1} - 1)}{N_{1}O_{1}}} \right]^{\text{T}}, \\
        \bm{a}_{2} &= \left[1, e^{j\frac{2\pi\theta_{2}}{N_{2}O_{2}}}, \ldots, e^{j\frac{2\pi\theta_{2} (N_{2} - 1)}{N_{2}O_{2}}} \right]^{\text{T}},
    \end{split}
\end{equation}
where $\theta_{1} \in \{0,1, \ldots, N_{1}O_{1}-1\}$ and $\theta_{2} \in \{0,1, \ldots, N_{2}O_{2}-1\}$ represent the beam indexes in horizontal and vertical dimensions, respectively.
$[\cdot]^{\text{T}}$ denotes the transpose.
Therefore, $\bm{a}_{1} \in \mathbb{C}^{N_{1} \times 1}$ and $\bm{a}_{2} \in \mathbb{C}^{N_{2} \times 1}$.

The supported configuration with respect to $(N_{1}, N_{2})$ and $(O_{1}, O_{2})$ can be found in \cite{38.214}.
Then, the set of spatial beams $\mathcal{B}$ based on 2D DFT is given by
\begin{equation}
    \mathcal{B} = \{\bm{b}_{\theta_{1},\theta_{2}}| \bm{b}_{\theta_{1},\theta_{2}} = \bm{a}_{1} \otimes \bm{a}_{2}\},
\end{equation}
where $\otimes$ indicates the Kronecker product, and $\bm{b}_{\theta_{1},\theta_{2}} \in \mathbb{C}^{N_{1}N_{2} \times 1}$.
Therefore, the whole candidate beam set contains $N_{1}N_{2}O_{1}O_{2}$ spatial beams.
In this paper, Type \uppercase\expandafter{\romannumeral1} codebook in codebookmode 1 is considered, where the optimal wideband beam is selected from the candidate beam set for the whole system bandwidth.

The antenna arrays of different polarizations employ the same spatial beam, but a co-phasing factor $\phi$ is used to quantize the phase difference between them.
Take the precoder of rank 4 in Type \uppercase\expandafter{\romannumeral1} codebook  when the number of transmit antenna is less than 16 as an example, which can be defined as
\begin{equation}
    \mathbf{W} = \frac{1}{\sqrt{4 N_{tx}}}
    \begin{bmatrix}
        \bm{b}_{\theta_{1},\theta_{2}}
        & \bm{b}_{\theta_{1}',\theta_{2}'} & \bm{b}_{\theta_{1},\theta_{2}}
        & \bm{b}_{\theta_{1}',\theta_{2}'} \\
        \phi \bm{b}_{\theta_{1},\theta_{2}}
        & \phi \bm{b}_{\theta_{1}',\theta_{2}'}
        & -\phi \bm{b}_{\theta_{1},\theta_{2}}
        & -\phi \bm{b}_{\theta_{1}',\theta_{2}'}
    \end{bmatrix},
\end{equation}
where the fraction is to normalize the precoder's power to 1 and the coefficient in the denominator has a 4 since $L = 4$.
$\phi \in \{1, j\}$, while a phase offset $\bm{\epsilon} = [\bm{\epsilon}_{1},\bm{\epsilon}_{2}]$ is set to horizontal and vertical beams.
Then, $\theta_{1}' = \theta_{1} + \bm{\epsilon}_{1}$ and $\theta_{2}' = \theta_{2} + \bm{\epsilon}_{2}$.
The values of $\bm{\epsilon}$ vary in different settings of $(N_{1}, N_{2})$ and $(O_{1}, O_{2})$, which can be obtained by looking up to \cite{38.214}.
Hence, $\mathbf{W} \in \mathbb{C}^{N_{tx} \times L}$ depends on four variables, i.e., $\theta_{1}$, $\theta_{2}$, $\bm{\epsilon}$ and $\phi$.
Each column in $\mathbf{W}$ is orthogonal to each other.

5G supports up to 8 layers for DL transmission, thus the set of Type \uppercase\expandafter{\romannumeral1} codebook $\mathcal{W}_{\text{codebook}}$ contains precoders from rank 1 to 8.
Clearly, more antenna ports are more likely to provide more throughput, since the beams are quantized with finer granularity to direct the energy toward the multipath of the channel.

The appropriate precoder can be chosen to maximize the sum mutual information over all subcarriers $\mathcal{K}$ and all layers $\mathcal{L}$ by an exhaustive search in codebook\cite{schwarz2010calculation}.
The mutual information when precoder $\mathbf{W}$ is employed can be expressed as
\begin{equation}
    I_{k,\ell}(\mathbf{W}) = \log(1+\text{SINR}_{k,\ell}(\mathbf{W})),
\end{equation}
where $k \in \mathcal{K}$, $\ell \in \mathcal{L}$, and $\text{SINR}_{k,\ell}(\mathbf{W})$ denotes the $\text{SINR}_{k,\ell}$ defined in (\ref{SINR}) if precoder $\mathbf{W}$ is adopted.
Then, the best precoder in the codebook at any time and location indexes is chosen as follows:
\begin{equation}
    \label{codebook-precoder}
    \begin{split}
        & \qquad \quad \mathbf{W}_{\text{PMI}^{*}} = \underset{\mathbf{W}_{\text{PMI}}}{\text{arg max}} \sum_{k \in \mathcal{K}} \sum_{\ell \in \mathcal{L}} I_{k,\ell}(\mathbf{W}_{\text{PMI}}) \\
        & \text{s.t.} \  \mathbf{W}_{\text{PMI}} \in \mathcal{W}_{\text{codebook}}, \
        \forall \ \text{PMI} \in \{1,\ldots, |\mathcal{W}_{\text{codebook}}|\},
    \end{split}
\end{equation}
where PMI indicates the index of the precoder in the codebook.
Then, the index of the chosen precoder is $\text{PMI}^{*}$, which is reported by UE to BS in 5G CLSM.
RI is the rank of $\mathbf{W}_{\text{PMI}^{*}}$, i.e., $L=r(\mathbf{W}_{\text{PMI}^{*}})$.
CQI is mapped from the effective SINR defined in (\ref{effSINR}) given the precoder and RI according to the CQI table in \cite{38.214}.
In 5G CLSM, with the help of CSI-RSs sent by BS, UE calculates the proper transmission parameters and reports $\{\text{PMI}^{*}, r(\mathbf{W}_{\text{PMI}^{*}}), \text{CQI}\}$ to BS for MIMO transmission.

\subsection{Statistic-based Solution}
In this subsection, we first analyze how to deal with channel data in time, frequency, and space domains to obtain the mapping, and then present the statistical solution based on 5G Type \uppercase\expandafter{\romannumeral1} codebook for data-driven MIMO.

In time domain, it is hard to infer transmission parameters since we do not utilize any channel feedback.
Fixed RI should be first determined, which indicates the rank of the precoder.
Then, we need to find the "representative" channel to get the precoder or obtain the "representative" precoder directly.
Since the precoder represents different spatial beams, we choose to fix it in time domain, which is more interpretable.
For CQI, it can be decided as the mode or mean value of the CQI set at one location according to the fluctuations of channels in time domain.

In frequency domain, channels on adjacent subcarriers will present correlation to some extent.
Thus, it is proper to handle the channels on grouped adjacent subcarriers by dividing $K$ channels into several groups, so as to obtain multiple sets of transmission parameters in frequency domain.
In this paper, we consider data transmission on narrow bands and employ one set of transmission parameters, i.e., the same precoder, RI, and CQI apply to all subcarriers.

In space domain, we need to infer the transmission parameters at unknown locations in $\mathcal{H}'$.
The fixed parameters in time and frequency domains derived from $\mathcal{H}$ facilitate the prediction in space domain, otherwise, we would obtain different inputs at the same location sample.
Also, the channels are somewhat correlated in space due to the similar signal propagation environment.
To determine the transmission parameters at unknown locations, RI should also be first chosen such that the fixed precoders derived from existing channel data can be utilized for inference.
Then, the integer interpolation method can be exploited to infer CQIs at unknown locations.

\begin{figure*}[t]
        \begin{center}
    \subfigure[Codebook-based approach]
    {
        \label{Codebook-based approach}
        \includegraphics[width=0.99\linewidth]{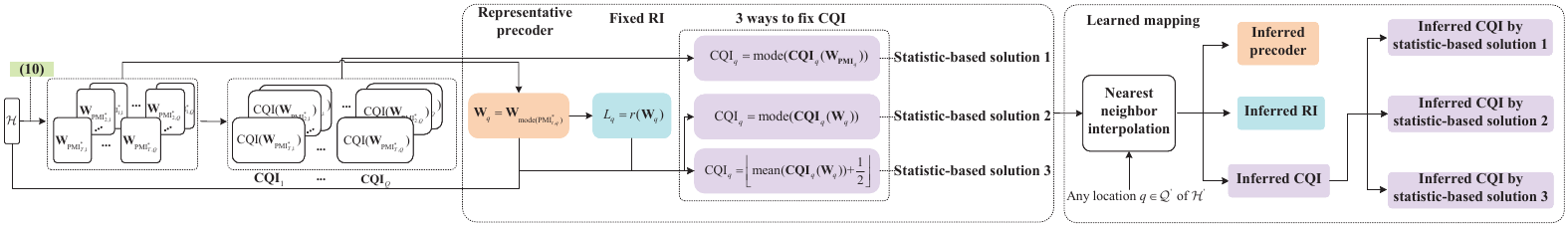}
    } \\
    \subfigure[SVD-based approach]
    {
        \label{SVD-based approach}
        \includegraphics[width=0.99\linewidth]{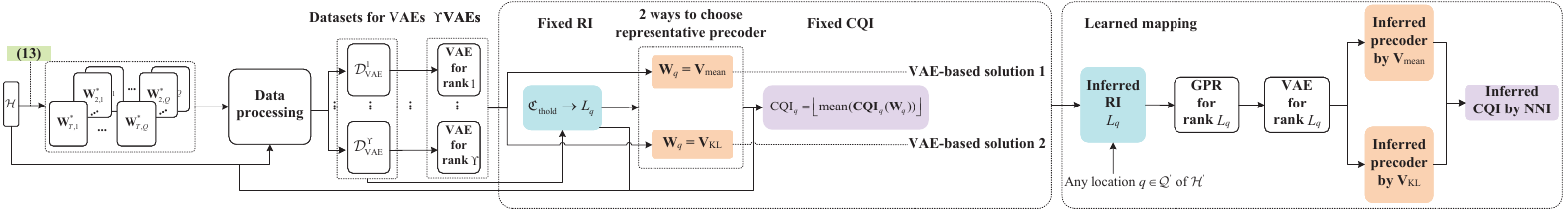}
    }
    \end{center}
    \caption{Flow charts of proposed approaches for data-driven MIMO.}
    \label{system flow chart}
\end{figure*} \par

For the codebook-based approach, the three transmission parameters are all quantized discrete numbers.
Therefore, the statistical solution can be used to fix transmission parameters as the values that occur most frequently from historical CSI based on the codebook of 5G, which is illustrated in Fig. \ref{Codebook-based approach}.
Denote $\textbf{PMI}_{q} = \{\text{PMI}^{*}_{t,q}| t \in \mathcal{T} \}$ as the historical PMI set at location $q$, the fixed precoder $\mathbf{W}_{q}$ is then chosen as $\mathbf{W}_{\text{mode}(\textbf{PMI}_{q})}$, where $\text{mode}(\cdot)$ indicates the mode value of the set.
$L_{q}$ is given by the rank of $\mathbf{W}_{q}$.
$\text{CQI}_{q}$ can also be determined by the mode of CQI set $\textbf{CQI}_{q}$.
$\textbf{CQI}_{q} = \textbf{CQI}_
{q}(\mathbf{W}_{\textbf{PMI}_{q}}) = \{\text{CQI}_{t,q}(\mathbf{W}_{\text{PMI}^{*}_{t,q}}) | t \in \mathcal{T}\}$ denotes the CQI set from CLSM.
Besides, the CQI set can be obtained from the results of fixed PMI and RI, which is denoted as $\textbf{CQI}_
{q}(\mathbf{W}_{q}) = \{\text{CQI}_{t,q}(\mathbf{W}_{q}) | t \in \mathcal{T}\}$.
The fixed CQI can then be decided as the mode of $\textbf{CQI}_
{q}(\mathbf{W}_{q})$ or the round of mean value of $\textbf{CQI}_
{q}(\mathbf{W}_{q})$ due to the little fluctuation of codebook-based approach.
Thus, different statistic-based solutions vary in the way to fix CQI.
As to spatial inference, nearest-neighbor interpolation can be utilized to predict transmission parameters at any location.

\section{SVD-based Approach}
\label{SA}
The quantization of spatial beams in the codebook brings in loss of accuracy of the precoder, yet the optimal precoder can be calculated from the SVD of the channel.
Therefore, we present another approach that starts by dealing with the optimal SVD precoders.
To capture the underlying meaning of high-dimensional precoders, we turn to VAE to find the representative one from the lower-dimensional latent representations.
Considering that the latent representations are Gaussian variables, we leverage GPR to generate new precoders for spatial prediction.
In this section, we first introduce how to obtain the optimal precoder from the SVD of the channel and then exploit VAE and GPR to implement data-driven MIMO based on SVD.

The optimal precoding matrix can be derived from SVD of the channel matrix for it eliminates interference among layers.
The SVD of channel matrix $\mathbf{H}_{k}$ can be expressed as
\begin{equation}
    \mathbf{H}_{k} = \mathbf{U}_{k}\mathbf{\Lambda}_{k}\mathbf{V}_{k}^{\dag},
\end{equation}
where $\mathbf{U}_{k} \in \mathbb{C}^{N_{rx} \times N_{rx}}$ are the left unitary matrix and $\mathbf{V}_{k} \in \mathbb{C}^{N_{tx} \times N_{tx}}$ are the right one.
$(\cdot)^{\dag}$ denotes the Hermitian transpose operation, which indicates the complex conjugate-transpose of the matrix.
The unitary matrix $\mathbf{V}_{k}$ satisfies the property that $\mathbf{V}_{k}\mathbf{V}_{k}^{\dag} = \mathbf{V}_{k}^{\dag}\mathbf{V}_{k} = \mathbf{I}_{N_{tx} }$.
Also, $\mathbf{U}_{k}\mathbf{U}_{k}^{\dag} = \mathbf{U}_{k}^{\dag}\mathbf{U}_{k} = \mathbf{I}_{N_{rx}}$.
$\mathbf{\Lambda}_{k} \in \mathbb{R}^{N_{rx} \times N_{tx}}$ is a rectangular matrix whose diagonal elements are non-negative real numbers, otherwise zeros.
Diagonal elements in $\mathbf{\Lambda}_{k}$ are named as singular values of $\mathbf{H}_{k}$, which are denoted as $\lambda_{i}$, $i \in \{1, \ldots, \Upsilon \}$.
Singular values are given in descending order, i.e., $\lambda_{1} \geqslant \lambda_{2} \geqslant \cdots \geqslant \lambda_{\Upsilon}$.

Then, the optimal precoder can be obtained as the first $L$ columns of $\mathbf{V}_{k}$, which is denoted as $\mathbf{V}_{k}^{L}$.
Meanwhile, if the selected equalizer is the first $L$ columns of $\mathbf{U}_{k}^{\dag}$, i.e., $\mathbf{E}_{k} = \mathbf{U}_{k}^{\dag, L}$, the post-equalization received symbol vector $\bm{m}_{k}$ in (\ref{post-equalization}) can be rewritten as
\begin{equation}
    \begin{split}
        \bm{m}_{k}
        & = \mathbf{U}_{k}^{\dag, L}\mathbf{H}_{k}\mathbf{V}_{k}^{L}\bm{s}_{k} + \mathbf{U}_{k}^{\dag, L}\bm{n}_{k} \\
        & = (\mathbf{U}_{k}^{\dag, L}\mathbf{U}_{k})\mathbf{\Lambda}_{k}(\mathbf{V}_{k}^{\dag}\mathbf{V}_{k}^{L})\bm{s}_{k} + \mathbf{U}_{k}^{\dag, L}\bm{n}_{k} \\
        & = \begin{bmatrix} \mathbf{I}_{L} & \mathbf{0} \end{bmatrix} \mathbf{\Lambda}_{k} \begin{bmatrix} \mathbf{I}_{L} \\ \mathbf{0} \end{bmatrix} \bm{s}_{k}
        + \mathbf{U}_{k}^{\dag, L}\bm{n}_{k} \\
        & = \mathbf{\Lambda}_{k}^{L,L} \bm{s}_{k} + \mathbf{U}_{k}^{\dag, L}\bm{n}_{k},
    \end{split}
\end{equation}
where $\mathbf{\Lambda}_{k}^{L,L}$ corresponds to the first $L$ rows and first $L$ columns of $\mathbf{\Lambda}_{k}$.
Hence, the MIMO transmission can be viewed as $L$ parallel SISO transmissions due to SVD of channel matrix.
With waterfilling power allocation, channel capacity can be achieved under certain constraints \cite{tse2005fundamentals}.

Similar to (\ref{codebook-precoder}), the best precoder from SVD of the channel at any time and location indexes is chosen as follows:
\begin{equation}
    \label{SVD-precoder}
    \begin{split}
        & \qquad \quad \mathbf{W}^{*} = \underset{\mathbf{W}_{i}}{\text{arg max}} \sum_{k \in \mathcal{K}} \sum_{\ell \in \mathcal{L}} I_{k,\ell}(\mathbf{W}_{i}) \\
        & \text{s.t.} \quad \mathbf{W}_{i} = \frac{\mathbf{V}_{k}^L}{\|\mathbf{V}_{k}^L\|_{F}} \in \mathcal{W}_{\text{SVD}},
        \ \forall \ i \in \{1,\ldots, |\mathcal{W}_{\text{SVD}}|\}, \\
        & \quad \quad \ \ \forall k \in \mathcal{K}, \ \forall L \in \{1,\ldots, \Upsilon \},
    \end{split}
\end{equation}
where $\mathcal{W}_{\text{SVD}}$ is the set of precoders obtained from the SVD of the channel at a certain location, and  $|\mathcal{W}_{\text{SVD}}| = K \times \Upsilon$.
$\mathbf{W}^{*}$ is the optimal precoder for the whole system bandwidth.
Note that $\mathbf{V}_{k}$ is a unitary matrix and its power equals to $L$.
Hence, it is divided by ${\|\mathbf{V}_{k}^L\|_{F}}$ to satisfy the power constraint in (\ref{problem}).
RI and CQI can then be determined as introduced before.

\subsection{VAE-based Solution}
\label{VAE-based Solution}

AE consists of an encoder and a decoder, which are neural networks of multiple layers.
The encoder compresses the initial input data and the decoder recovers them from lower-dimensional latent space.
However, traditional AE fails to generate new meaningful contents, since inputs are encoded into discrete points, leading to the irregularity of latent space.
Therefore, VAE is firstly introduced in \cite{kingma2013auto} to make the latent space continuous.
Instead of directly mapping data into points, VAE's encoder returns a distribution over the latent space.
Then, the decoder uses points randomly sampled from latent space to recover the initial data.
Since the latent space of VAE is regular, we can sample any point from it to generate new data via decoder, which fits well for our purpose of understanding the high dimensional precoders and inferring them in space domain.

Specifically, $\bm{d}_i$ is denoted as the $i$-th independent and identically distributed sample of dataset $\mathcal{D}_{\text{VAE}}$ for VAE, where $i \in \{ 1, \ldots, |\mathcal{D}_{\text{VAE}}| \}$.
It is assumed that outputs of VAE are generated by a random process, corresponding to the latent variable $\bm{\varpi}$ with dimension $N_{lv} \times 1$.
The encoder network's parameters are denoted as $\bm{\psi}$, and the decoder network's as $\bm{\xi}$.
$\bm{\psi}$ and $\bm{\xi}$ are unknown parameters, which require to be learned from $\mathcal{D}_{\text{VAE}}$.
The objective is to maximize the marginal likelihood distribution $p_{\bm{\xi}}(\bm{d})$ such that the generative process can accurately recover the inputs.

Nevertheless, the integral of marginal likelihood has no closed-form solution.
Thus, the variational inference is exploited in VAE to handle the intractability of the marginal likelihood by introducing the variational distribution $\widehat{p}_{\bm{\psi}}(\bm{\varpi}|\bm{d})$ to approximate the true posterior distribution $p_{\bm{\xi}}(\bm{\varpi}|\bm{d})$ \cite{kingma2013auto}.
Then, the evidence lower bound (ELBO) for the log-likelihood distribution $\log p_{\bm{\xi}}(\bm{d}_{i})$ can be derived as
\begin{equation}
    \label{ELBO}
    \begin{split}
       \text{ELBO}(\bm{\psi},\bm{\xi},\bm{d}_{i})
       & = \mathbb{E}_{\widehat{p}_{\bm{\psi}}(\bm{\varpi}|\bm{d}_{i})}\left[\log p_{\bm{\xi}}(\bm{d}_{i}|\bm{\varpi})\right] \\
       & - \text{KL}(\widehat{p}_{\bm{\psi}}(\bm{\varpi}|\bm{d}_{i})||p_{\bm{\xi}}(\bm{\varpi})),
    \end{split}
\end{equation}
where KL is the Kullback–Leibler divergence between two distributions, which measures the difference between them in terms of distribution.
The ELBO depends on the parameters of both decoder and encoder networks, along with the input data point.
The marginal log-likelihood $\log p_{\bm{\xi}}(\bm{d}_{i}, \ldots, \bm{d}_{|\mathcal{D}_{\text{VAE}}|}) = \sum_{i=1}^{|\mathcal{D}_{\text{VAE}}|} \log p_{\bm{\xi}}(\bm{d}_{i})$.
Then, maximizing the marginal log-likelihood is equivalent to maximizing the ELBO.

In VAE, the prior $p_{\bm{\xi}}(\bm{\varpi})$ and posterior approximation $\widehat{p}_{\bm{\psi}}(\bm{\varpi}|\bm{d}_{i})$ are both assumed to be Gaussian distribution.
$p_{\bm{\xi}}(\bm{\varpi}) \sim \mathcal{N}(\bm{0},\mathbf{I})$ is a standard multivariate Gaussian distribution, and $\widehat{p}_{\bm{\psi}}(\bm{\varpi}|\bm{d}_{i}) \sim \mathcal{N}(\bm{\mu}(\bm{\psi},\bm{d}_{i}),\bm{\sigma}^2(\bm{\psi},\bm{d}_{i})\mathbf{I})$ is a multivariate Gaussian distribution with mean vector $\bm{\mu}(\bm{\psi},\bm{d}_{i})$ and variance vector $\bm{\sigma}^2(\bm{\psi},\bm{d}_{i})$.
Thus, the encoder maps the input $\bm{d}_{i}$ into vectors of mean and variance to obtain the posterior approximation  $\widehat{p}_{\bm{\psi}}(\bm{\varpi}|\bm{d}_{i})$.
Since the sampling operation of latent space lacks gradient information, the reparameterization trick is used in VAE to make gradient decent possible, which can be expressed as
\begin{equation}
    \label{reparameterization}
    \Tilde{\bm{\varpi}}(\bm{\psi},\bm{d}_{i}) = \bm{\mu}(\bm{\psi},\bm{d}_{i}) + \bm{\sigma}(\bm{\psi},\bm{d}_{i}) \odot \bm{\varepsilon},
\end{equation}
where $\bm{\varepsilon} \sim \mathcal{N}(\bm{0},\mathbf{I})$ and $\odot$ denotes the element-wise product.
The sampled latent variable $\Tilde{\bm{\varpi}}(\bm{\psi},\bm{d}_{i})$ is the function of parameters of encoder and input data point, so are $\bm{\mu}(\bm{\psi},\bm{d}_{i})$ and $\bm{\sigma}(\bm{\psi},\bm{d}_{i})$.
For simplicity, we leave out $(\bm{\psi},\bm{d}_{i})$ in the following part when referring to these three variables.

We denote the output of VAE as $\widehat{\bm{d}_{i}}, i \in \{ 1, \ldots, |\mathcal{D}_{\text{VAE}}| \}$, which can be viewed as the estimate of input $\bm{d}_{i}$.
The negative ELBO can be chosen as the loss function, which is minimized during the training process of VAE and can be defined as
\begin{equation}
    \label{loss}
    \begin{split}
       \text{Loss}(\bm{\psi},\bm{\xi},\bm{d}_{i})
       & = \text{SE}(\bm{d}_{i}, \widehat{\bm{d}_{i}}) \\
       & + \beta \text{KL}
       \left( \mathcal{N}(\bm{\mu},\bm{\sigma}^2\mathbf{I})||\mathcal{N}(\bm{0},\mathbf{I}) \right) \\
       & = ||\bm{d}_{i} - \widehat{\bm{d}_{i}}||^2 \\
       & + \frac{\beta}{2} \sum_{j=1}^{N_{lv}}
       \left(\bm{\mu}_{j}^2 + \bm{\sigma}_{j}^2 - \log(\bm{\sigma}_{j}^2) - 1\right),
    \end{split}
\end{equation}
where the first term indicates the reconstruction loss in squared error form.
The second term refers to the KL divergence loss between the posterior approximation and the prior of latent variable, which is integrated analytically in the Gaussian case.
To minimize the loss function is to reduce the reconstruction error and keep the posterior approximation close to standard Gaussian distribution at the same time, where the latter can be considered as a regularisation term to the latent space.
$\beta$ is an adjustable hyperparameter introduced in \cite{higgins2016beta} to balance the tradeoff between reconstruction loss and KL divergence loss.
During the training process, a minibatch of input data with size $N_{b}$ is used to calculate the loss, which will be further utilized to perform gradient descent to update the parameters of networks of encoder and decoder by an optimization algorithm.

\begin{figure*}[t]
    \begin{center}
    \subfigure[VAE architecture for precoders of rank 4]
    {
        \includegraphics[width=0.95\linewidth]{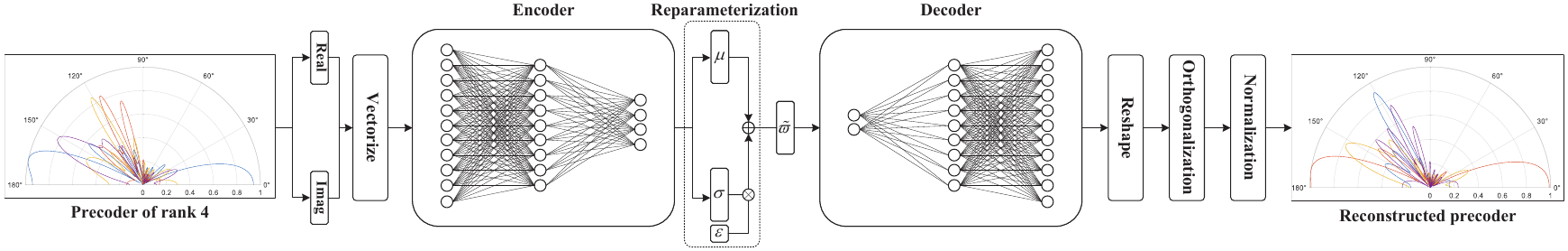}
    } \\
    \subfigure[Fixing precoder of rank 4 in time domain]
    {
        \includegraphics[width=0.55\linewidth]{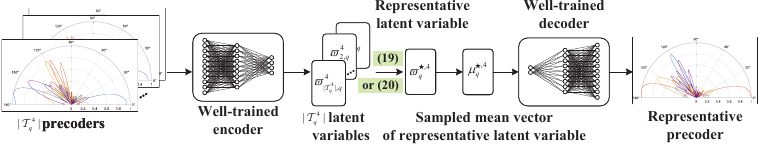}
    }
    \subfigure[Inferring precoder of rank 4 in space domain]
    {
        \includegraphics[width=0.38\linewidth]{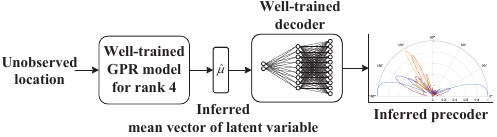}
    }
    \end{center}
    \caption{Our proposed VAE-based solution to dealing with optimal precoders of rank 4 obtained from SVD of channels.}
    \label{VAE_illustration}
\end{figure*}

In the following parts of this subsection, we introduce how to implement the SVD-based approach leveraging VAE.
Fig. \ref{SVD-based approach} illustrates the flow chart of the SVD-based approach.
We first describe how to generate the dataset for VAE.
Then, RI is fixed according to a percentage count threshold.
We propose two methods to choose the representative precoders at different locations in $\mathcal{H}$, and fixed CQI can be determined from the CQI set of fixed precoder and RI.
At last, we utilize GPR to infer the precoder at any location in $\mathcal{H}'$ and exploit NNI for CQI interpolation.
The details of how we utilize VAE for precoders of rank 4 based on SVD are demonstrated in Fig. \ref{VAE_illustration}.

 \par

\subsubsection{Fixing of Transmission Parameters}

\label{Fixing of Transmission Parameters}

In this paper, we use VAE to reduce the dimension of the precoder to obtain its underlying characteristics, which are reflected by the latent variable.
Then, the representative precoder can be derived at different locations by choosing the representative latent variable and inputting it into the well-trained decoder network.
Since precoders of different layers have different dimensions, $\Upsilon$ VAEs are required to be trained corresponding to distinct input dimensions.

However, how to prepare the input precoder dataset remains a problem.
Basically, we can obtain the optimal precoders $\mathbf{W}^{*}_{t,q}$ at time $t$ and location $q$ from existing channel data according to (\ref{SVD-precoder}) and $r(\mathbf{W}^{*}_{t,q})$ varies at different time and location indexes.
The choice of precoder of different layers possesses downward compatibility, which means that the precoder of lower rank can also be used even if the optimal precoder is of higher rank.
In this way, the dataset of precoders of lower rank has more elements.
If we fix the rank of precoder $L$ obtained from SVD of channel to $\mathfrak{r}$ in (\ref{SVD-precoder}), we can get "optimal" precoders of rank $\mathfrak{r}$, which are denoted as $\mathbf{W}^{*, \mathfrak{r}}$.
Then, the chosen indexes of time and location for precoders with higher rank are also applicable to precoders with lower rank.
The downward compatible set of layer choice for precoders of rank $\mathfrak{r}$ is denoted as $\mathfrak{L}$, where $\mathfrak{L} = \{\mathfrak{r}, \ldots, \Upsilon\}$ and $\mathfrak{r} \in \{1, \ldots, \Upsilon\}$.

Specifically, the indexes of time and location corresponding to the selected precoders $\mathbf{W}^{*, \mathfrak{r}}$ can be obtained through constraints on the original optimal precoders $\mathbf{W}^{*}$.
First, the location index set $\mathcal{Q}^{\mathfrak{r}}$ for precoders of rank $\mathfrak{r}$ is chosen as $\mathcal{Q}^{\mathfrak{r}} = \{q | \text{mode}(\mathcal{R}_{q}(\mathbf{W}^{*}_{\mathcal{T},q})) \in \mathfrak{L}, q \in \mathcal{Q}\}$,
where $\mathcal{R}_{q}(\mathbf{W}^{*}_{\mathcal{T},q}) = \{r(\mathbf{W}^{*}_{t,q}) | t \in \mathcal{T} \}$ is the rank set of optimal precoders at location index $q$.
Then, time index set $\mathcal{T}^{\mathfrak{r}}_{q}$ at location $q$ for precoders of rank $\mathfrak{r}$ is selected as $\mathcal{T}^{\mathfrak{r}}_{q} = \{t |
    r(\mathbf{W}^{*}_{t,q}) \in \mathfrak{L}, t \in \mathcal{T}  \},
    \forall q \in \mathcal{Q}^{\mathfrak{r}}$.
$\mathcal{T}^{\mathfrak{r}}_{q}$ differs among different location indexes.
Evidently, $\mathcal{Q}^{\mathfrak{r}} \subseteq \mathcal{Q}$ and
$\mathcal{T}^{\mathfrak{r}}_{q} \subseteq \mathcal{T}, \forall q \in \mathcal{Q}^{\mathfrak{r}}$.
Finally, the dataset of VAE $\mathcal{D}_{\text{VAE}}^{\mathfrak{r}}$ for precoders of rank $\mathfrak{r}$ can be expressed as $\mathcal{D}_{\text{VAE}}^{\mathfrak{r}} = \{ \mathbf{W}^{*, \mathfrak{r}}_{t,q}| q \in \mathcal{Q}^{\mathfrak{r}}, t \in \mathcal{T}^{\mathfrak{r}}_{q}\}$.

Note that the values of $\mathbf{W}^{*}$ are small due to the power normalization, which makes it hard for VAE to learn the underlying characteristics.
Hence, the optimal precoder without power normalization $\mathbf{V}^{*, L}$ is directly used in the dataset to replace $\mathbf{W}^{*}$.
$\mathbf{V}^{*, L}$ is then reshaped to a vector of size $N_{tx} \times L \times 2 $, where 2 refers to the real and imaginary parts of complex-valued precoder.
The outputs of VAE are then reshaped into the original size of $\mathbf{V}^{*, L}$.
Besides, an orthogonalization operation is performed to the recovered precoder $\widehat{\mathbf{V}}^{*, L}$ to make it satisfy the property of the unitary matrix.
The orthogonalization operation is done by QR decomposition of $\widehat{\mathbf{V}}^{*,L}$ \cite{golub2013matrix}, which is given by: $\widehat{\mathbf{V}}^{*,L} = \widehat{\mathbf{V}}^{*,L}_{\mathbf{Q}} \widehat{\mathbf{V}}^{*,L}_{\mathbf{R}}$.
$\widehat{\mathbf{V}}^{*,L}_{\mathbf{Q}} \in \mathbb{C}^{N_{tx} \times L}$ is an orthogonal matrix with columns as unit vectors and $\widehat{\mathbf{V}}^{*,L}_{\mathbf{R}} \in \mathbb{C}^{L \times L}$ is an upper triangular matrix.
 Finally, $\widehat{\mathbf{V}}^{*, L}_{\mathbf{Q}} / \|\widehat{\mathbf{V}}^{*, L}_{\mathbf{Q}}\|_{F}$ is decided as the generated or recovered precoder, with power normalized and columns orthogonalized.

In this paper, the existing channel data $\mathcal{H}$ is the training dataset, from which the precoder dataset is derived.
We fix transmission parameters at locations in $\mathcal{H}$ to facilitate the inference at any location.
We need to first fix RI such that the representative precoder of the corresponding rank can be chosen.
Specifically, we define a percentage count threshold $\mathfrak{C}_{\text{thold}}$.
If no fewer than $\mathfrak{C}_{\text{thold}}$ percents of $T$ RIs obtained from (\ref{SVD-precoder}) at location $q$ are elements in $\mathfrak{L}$ when the mode of $\mathcal{R}_{q}(\mathbf{W}^{*}_{\mathcal{T},q})$ is $\mathfrak{r}$, then $L_{q}$ is fixed to $\mathfrak{r}$, otherwise $\mathfrak{r}-1$.
Thus, apart from locations where the highest rank of precoder is employed, the set of location $\mathcal{Q}^{\mathfrak{r}, \mathfrak{C}_{\text{thold}}}_{\text{fixed}}$ where the fixed RI is $\mathfrak{r}$ consists of two parts: the indexes that satisfy the constraints described above when the mode of $\mathcal{R}_{q}(\mathbf{W}^{*}_{\mathcal{T},q})$ is $\mathfrak{r}$ and those that don't when the mode of $\mathcal{R}_{q}(\mathbf{W}^{*}_{\mathcal{T},q})$ is $\mathfrak{r}+1$.
$\mathcal{Q}^{\mathfrak{r}, \mathfrak{C}_{\text{thold}}}_{\text{fixed}} \subseteq \mathcal{Q}^{\mathfrak{r}}$ and $\sum_{\mathfrak{r}=1}^{\Upsilon} |\mathcal{Q}^{\mathfrak{r}, \mathfrak{C}_{\text{thold}}}_{\text{fixed}}| = |\mathcal{Q}| = Q$.
A higher value of $\mathfrak{C}_{\text{thold}}$ indicates a more conservative choice of fixed RI.

The precoder at location $q$ can be fixed, exploiting the lower dimension latent variables obtained by the encoder of VAE.
Specifically, the representative latent variable $\bm{\varpi}_{q}^{\star, \mathfrak{r}}$ at location $q$ for precoders of rank $\mathfrak{r}$ can be chosen as the one $\bm{\varpi}^{\mathfrak{r}}_{t^{\star}(q),q}$ with the representative time index $t^{\star, {\mathfrak{r}}}(q)$, which has the minimum distance to the mean value of mean vectors $\bm{\mu}_{t,q}^{\mathfrak{r}}$ and variance vectors $(\bm{\sigma}_{t,q}^{\mathfrak{r}})^2$ of all latent variables at location $q$.
Then, $t^{\star, {\mathfrak{r}}}(q)$ is selected as
\begin{equation}
    \begin{aligned}
       \underset{t}{\text{arg min}}
    &[ (\bm{\mu}_{t,q}^{\mathfrak{r}} - \Bar{\bm{\mu}}_{q}^{\mathfrak{r}})^\text{T}(\bm{\mu}_{t,q}^{\mathfrak{r}} - \Bar{\bm{\mu}}_{q}^{\mathfrak{r}}) +  \\
    &  ((\bm{\sigma}_{t,q}^{\mathfrak{r}})^2 - (\Bar{\bm{\sigma}}_{q}^{\mathfrak{r}})^2)^\text{T}((\bm{\sigma}_{t,q}^{\mathfrak{r}})^2 - (\Bar{\bm{\sigma}}_{q}^{\mathfrak{r}})^2) ], \quad \forall t \in \mathcal{T}_{q}^{\mathfrak{r}},
    \end{aligned}
 \end{equation}
where $\Bar{\bm{\mu}}_{q}^{\mathfrak{r}} = \sum_{t \in \mathcal{T}_{q}^{\mathfrak{r}}} \bm{\mu}_{t,q}^{\mathfrak{r}} / |\mathcal{T}_{q}^{\mathfrak{r}}|$ is the mean value of mean vectors, and $(\Bar{\bm{\sigma}}_{q}^{\mathfrak{r}})^2 = \sum_{t \in \mathcal{T}_{q}} (\bm{\sigma}_{t,q}^{\mathfrak{r}})^2 / |\mathcal{T}_{q}^{\mathfrak{r}}|$ is that of variance vectors.
Then, we take the mean vector $\bm{\mu}_{q}^{\star, \mathfrak{r}}$ of $\bm{\varpi}_{q}^{\star,\mathfrak{r}}$ as its sampled values, since the probability of obtaining the mean of a Gaussian variable is highest.
Finally, we input $\bm{\mu}_{q}^{\star,\mathfrak{r}}$ into the well-trained decoder of VAE to obtain the representative precoder.
Since this method is based on finding the latent variable closest to the statistical mean value, we name the recovered or generated precoder as $\mathbf{V}_{\text{mean}}$.

Another way to choose the representative latent variable is to find the one closest to others in terms of distribution. The chosen time index $t^{\star,\mathfrak{r}}(q)$ can be defined as
\begin{equation}
         \underset{t}{\text{arg min}}
    \sum_{t' \in \mathcal{T}_q^{\mathfrak{r}} \setminus t}
    \left\{
    \begin{aligned}
        & \text{KL}(D(\bm{\varpi}_{t,q}^{\mathfrak{r}})||D(\bm{\varpi}_{t',q}^{\mathfrak{r}}))
        , \text{if } t < t'
        \\
        & \text{KL}(D(\bm{\varpi}_{t',q}^{\mathfrak{r}})||D(\bm{\varpi}_{t,q}^{\mathfrak{r}})),
        \text{if } t' < t
    \end{aligned}
    \right. , \forall t \in \mathcal{T}_{q}^{\mathfrak{r}},
 \end{equation}
where $D(\cdot)$ denotes the distribution of the variable.
There are two cases to make sure that the KL divergence is calculated for the distribution of lower-indexed latent variable from higher-indexed since KL divergence is not symmetric in the two distributions.
Leaving out subscript $q$, $\text{KL}(D(\bm{\varpi}_{t}^{\mathfrak{r}})||D(\bm{\varpi}_{t'}^{\mathfrak{r}}))$ in Gaussian case is given by
\begin{equation}
    \begin{aligned}
     & \text{KL} \left( \mathcal{N}(\bm{\mu}_{t}^{\mathfrak{r}},(\bm{\sigma}^{\mathfrak{r}}_{t})^2\mathbf{I})||\mathcal{N}(\bm{\mu}_{t'}^{\mathfrak{r}},(\bm{\sigma}_{t'}^{\mathfrak{r}})^2\mathbf{I}) \right) =
     \frac{1}{2} \sum_{j=1}^{N_{lv}} \bigg(\log(\bm{\sigma}_{t',j}^{\mathfrak{r}})^2
     \\
    &  +
       \frac{(\bm{\mu}_{t,j}^{\mathfrak{r}} - \bm{\mu}_{t',j}^{\mathfrak{r}})^2 + (\bm{\sigma}_{t,j}^{\mathfrak{r}})^2}
      {(\bm{\sigma}_{t',j}^{\mathfrak{r}})^2}
      - \log(\bm{\sigma}_{t,j}^{\mathfrak{r}})^2 - 1\bigg) ,
    \end{aligned}
\end{equation}
where $(\cdot)_{t,j}$ refers to the $j$-th index of variable at time $t$ and any location, which replaces $(\cdot)_{t,q,j}$.
Since this method chooses the representative latent variable according to KL divergence, we name the output precoder as $\mathbf{V}_{\text{KL}}$.
At location $q$, the CQI set $\textbf{CQI}_
{q}(\mathbf{W}_{q})$ can be determined when the fixed RI and representative precoder obtained from VAE are employed.
Then, the fixed CQI is decided as the floor of the mean value of $\textbf{CQI}_
{q}(\mathbf{W}_{q})$ since fixed SVD precoder may lead to more fluctuations in time domain.

\subsubsection{Inference of Transmission Parameters}
In data-driven MIMO, the performance of fixed transmission parameters is first evaluated on $\mathcal{H}$.
Then, the performance of inference on transmission parameters at any location is tested on $\mathcal{H}'$.
Thus, $\mathcal{H}'$ can be viewed as the testing dataset.

RI should be first determined such that the representative precoders of different layers can be utilized for prediction at any location.
Since higher RI would lead to more interference among data streams, we have a conservative choice of RI and determine the value of RI at any location in $\mathcal{H}'$ as the minimal value of fixed RIs at $N_{\text{RI}}$ closest locations in $\mathcal{H}$.
Since the latent representations are multivariate Gaussian variables, we use GPR to predict the latent variable at any location and then feed it into the decoder of VAE to generate the inferred precoder.

GPR is a non-parameter regression model based on probability \cite{rasmussen2006gaussian}.
It assumes that input data are the samples of the Gaussian process, which obey the Gaussian distribution and so do the latent variables in VAE.
In this paper, similar to VAE, we need to train $\Upsilon$ GPRs.
Consider a dataset $\mathcal{D}_{\text{GPR}}^{\mathfrak{r}} = \{ (\bm{\varrho}_{q}, \bm{\mu}_{q}^{\star, \mathfrak{r}}) | q \in \mathcal{Q}^{\mathfrak{r}} \}$ for GPR to infer precoders of rank $\mathfrak{r}$, where $\bm{\mu}_{q}^{\star,\mathfrak{r}}$ is the mean vector of the representative latent variable and is also the desirable output corresponding to input location $\bm{\varrho}_{q}$.
GPR fits the unknown function by utilizing a kernel function $\mathfrak{K}$, which measures the similarity between two input locations.
Close inputs are expected to have similar outputs in GPR.
The most common kernel function is radial basic function (RBF), which is defined as
\begin{equation}
    \mathfrak{K}(\bm{\varrho}_{q}, \bm{\varrho}_{q'}) =
    \gamma^2 e^{-\frac{(\bm{\varrho}_{q} - \bm{\varrho}_{q'})(\bm{\varrho}_{q} - \bm{\varrho}_{q'})^{\text{T}}}{2\zeta^2}},
\end{equation}
where $\gamma$ and $\zeta$ are hyperparameters concerning width and characteristic length-scale, respectively.

Let $\bm{\Gamma}$ be the combined observed location matrix with each column as $\bm{\varrho}_{q}$, and $\bm{\Delta}$ be the combined known output matrix with each column as $\bm{\mu}_{q}^{\star, \mathfrak{r}}$, where $q \in \mathcal{Q}^{\mathfrak{r}}$.
The regression model $p(\bm{\Delta}|\bm{\Gamma})$ is assumed to be a multivariate Gaussian distribution with the mean vector of zeros and covariance matrix $\bm{\Xi}(\bm{\Gamma}, \bm{\Gamma})$.
$\bm{\Xi}(\bm{\Gamma}, \bm{\Gamma}) \in \mathbb{R}^{|\mathcal{Q}^{\mathfrak{r}}| \times |\mathcal{Q}^{\mathfrak{r}}|}$ characterises the correlation between observed location points $\bm{\Gamma}$, and its element in row $i$ and column $j$ is $\mathfrak{K}(\bm{\varrho}_{i}, \bm{\varrho}_{j})$.
Then, $p(\bm{\Delta}|\bm{\Gamma})$ is maximized during the training process, and optimal hyperparameters of RBF can be obtained.
Based on the properties of multivariate Gaussian, the posterior probability $p(\hat{\bm{\Delta}}|\bm{\Delta})$ is maximized when the predicted mean vectors $\hat{\bm{\Delta}}$ of latent variables at unobserved locations $\hat{\bm{\Gamma}}$ equal to $\bm{\Xi}(\hat{\bm{\Gamma}}, \bm{\Gamma})\bm{\Xi}(\bm{\Gamma}, \bm{\Gamma})^{-1}\bm{\Delta}$.
$\bm{\Xi}(\hat{\bm{\Gamma}}, \bm{\Gamma})$ describes the correlation between unobserved locations $\hat{\bm{\Gamma}}$ and observed ones $\bm{\Gamma}$, while $(\cdot)^{-1}$ indicates the inverse matrix.
By inputting $\hat{\bm{\Delta}}$ to the well-trained decoder network of VAE, we can obtain the inferred precoders at unobserved locations.

As to CQI, NNI can be used as the integer interpolation method to obtain interpolated CQI at any location in $\mathcal{H}'$ based on the fixed CQI at locations in $\mathcal{H}$.
NNI is a spatial interpolation method based on the Voronoi diagram \cite{amidror2002scattered}, which is a partition of a space obtained from the perpendicular bisectors of points given the known sample set.
Specifically, we can obtain the original Voronoi diagram based on fixed CQIs at locations in $\mathcal{H}$.
New location point $q'$ in $\mathcal{H}'$ will lead to the change in the Voronoi cells of points close to it, then interpolated CQI at location $q'$ is given by
\begin{equation}
    \text{CQI}_{q'} = \sum_{q \in \mathcal{Q}(q')} \frac{\mathfrak{V}_{q} \cap \mathfrak{V}_{q'}}{\mathfrak{V}_{q'}} \text{CQI}_{q}, \quad \forall q' \in \mathcal{Q}',
\end{equation}
where $\mathcal{Q}(q')$ is the set of original space indexes that are close to new location sample $q'$ and $\text{CQI}_{q}$ is the fixed CQI at location $q$.
The volume of new Voronoi cell centered at location $q'$ is denoted as $\mathfrak{V}_{q'}$, and the intersected volume between the original cell centered at $q$ and the new cell centered at $q$' is denoted as $\mathfrak{V}_{q} \cap \mathfrak{V}_{q'}$.
Therefore, $\mathfrak{V}_{q} \cap \mathfrak{V}_{q'} / \mathfrak{V}_{q'}$ serves as the interpolation weight, where $0 < \mathfrak{V}_{q} \cap \mathfrak{V}_{q'} / \mathfrak{V}_{q'} \leqslant 1$ and $\sum_{q \in \mathcal{Q}(q')}\mathfrak{V}_{q} \cap \mathfrak{V}_{q'} / \mathfrak{V}_{q'} = 1, \forall q' \in \mathcal{Q}'$.

\section{Simulation Results and Discussions}
In this section, we introduce the simulation setup of data-driven MIMO and present the simulation results and discussions of proposed solutions for two different approaches.

\label{SR and D}

\subsection{Simulation Setup}
\subsubsection{Environment}

In this paper, throughput is obtained by signal processing in the physical layer based on the Vienna 5G link level simulator \cite{pratschner2018versatile}.
BS is equipped with $N_{tx} = 16$ dual polarized transmit antennas with configuration $(N_{1},N_{2}) = (8,1)$ and UE is equipped with $N_{rx} = 4$ single polarized receive antennas.
$\Upsilon$ is given by $\min\{N_{tx}, N_{rx}\} = 4$.
A total of $K = 72$ subcarriers are used to carry MIMO transmission for UE.
The center frequency is 3.5 GHz and the subcarrier spacing is 15 kHz.
Low-density parity-check (LDPC) channel coding is employed and the equalizer is chosen based on minimum mean square error (MMSE).

\begin{figure}[t]
    \begin{center}
        \includegraphics[width=0.45\textwidth]{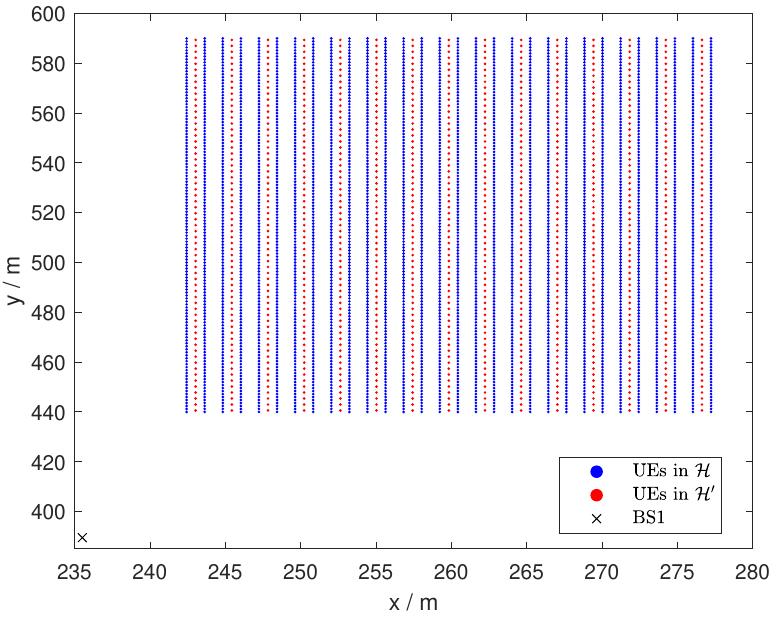}
        \caption{Locations of selected UEs and BS in DeepMIMO \emph{O1} scenario in the $x$ and $y$ directions.}
        \label{locations}
    \end{center}
\end{figure}

Meanwhile, simulations are conducted based on the realistic ray-tracing dataset: DeepMIMO \cite{alkhateeb2019deepmimo}.
Using accurate ray-tracing data obtained from Remcom Wireless InSite, cluster delay line (CDL) channels between BS and UE are generated according to standard \cite{38.901}, which serve as the dataset in this paper.
\emph{O1} is the chosen outdoor scenario of DeepMIMO.
The DeepMIMO dataset has many candidate BSs and UEs with different coordinates to choose from, while all BSs' height is 6m and that of UEs is 2m.
In this paper, we focus on using \emph{BS1} to transmit data to UEs in \emph{User Grid 1} in \emph{O1} scenario of DeepMIMO.
The training set consists of channels generated between \emph{BS1} and UEs of 126 rows from \emph{R715} to \emph{R1465} in \emph{O1} with equal spacing, where each row contains 30 user locations.
Thus, $|\mathcal{Q}| = 3780$.
Similarly, the testing set is made up of channels generated between \emph{BS1} and UEs of 63 rows from \emph{R718} to \emph{R1462} in \emph{O1} with equal spacing, where each row contains 15 user locations.
Then, $|\mathcal{Q}'| = 945$, which corresponds to the 80-20 split of training and testing sets.
Besides, 100 different samples of channels in time domain are considered.
Detailed locations of selected UEs and BS are shown in Fig. \ref{locations}.

 \par


\subsubsection{VAE}
Although a higher accuracy of the recovered precoder can be obtained if the network in VAE gets deeper, we found that the loss wouldn't reduce too much after more than 3 hidden layers are employed for both encoder and decoder networks.
Hence, in this paper, they are both designed as fully connected neural networks with 3 hidden layers.
For the encoder, the input dimension is $N_{tx} \times L \times 2 = 32 L$.
The 3 hidden layers for the encoder are of size 400, 128, and $ N_{lv} \times 2$, where 2 indicates the means and log variances of the latent variable.
$N_{lv}$ is set to $10L$.
In this paper, we output the log variance of the latent variable to ensure the variance is always positive.
Leaky rectified linear unit (ReLU) is chosen to be the activation function for the encoder.
For the decoder, the input dimension is $N_{lv}$.
Conversely, the 3 hidden layers for the decoder are of size 128, 400, and $32 L$.
The activation functions for the decoder are also leaky ReLU, except that the last layer is Tanh to obtain the outputs whose values are between -1 and 1.

During the training process, a minibatch of $N_{b} = 128$ samples is used to calculate the gradients.
$\beta$ in (\ref{loss}) is chosen to be 0.01.
Then, the parameters of networks in VAE are updated for 100 epochs using ADAM \cite{kingma2014adam} optimization algorithm with $10^{-3}$ learning rate.
Despite fine-tuning may be helpful, we found that the proposed VAE solution works well under many values of hyperparameters.

\subsection{Evaluation on Time and Frequency Domains}
In this subsection, we first compare the performance loss of the codebook-based approach when transmitting single and multiple streams, respectively.
Then, we verify the function of the previously mentioned tricks on VAE at samples of location suitable for transmitting four data streams and extend to all location points in $\mathcal{H}$.
Here, samples suitable for transmitting four streams refer to location points that utilize precoders of rank 4 at all time indexes based on the results of SVD obtained from (\ref{SVD-precoder}), i.e., $\mathcal{Q}^{4,100}_{\text{fixed}}$.

\subsubsection{Single Stream and Multiple Streams Transmission}

\begin{figure}[htbp]
    \begin{center}
    \subfigure[single stream, $N_{rx}=1$]
    {
        \label{single stream}
        \includegraphics[width=0.45\linewidth]{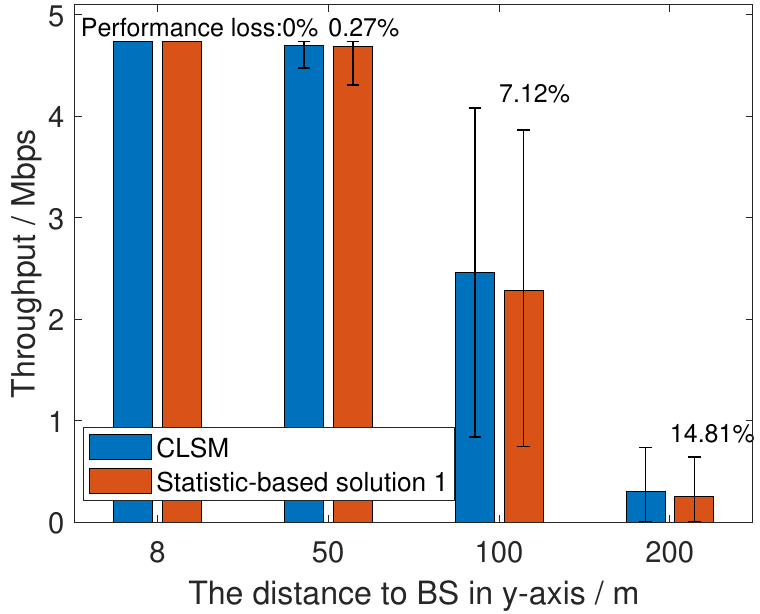}
    }
    \subfigure[multiple streams, $N_{rx}=4$]
    {
        \label{multiple streams}
        \includegraphics[width=0.45\linewidth]{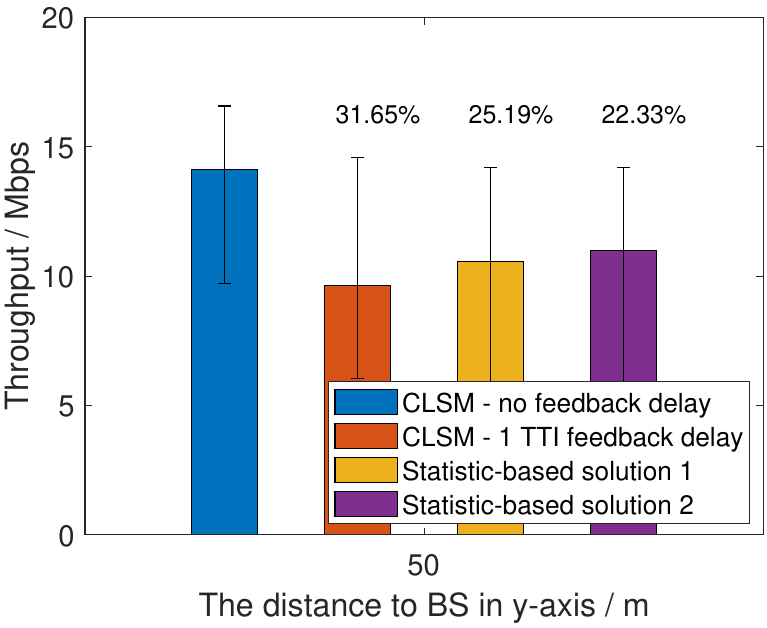}
    }
    \end{center}
    \caption{Average throughput comparison between CLSM and statistic-based solutions when transmitting single and multiple streams.}
    \label{single stream vs. multiple streams}
\end{figure} \par

Fig. \ref{single stream vs. multiple streams} shows the average throughput comparison between CLSM and statistic-based solutions when transmitting single and multiple streams.
Throughput is first averaged in time and frequency domains and then is averaged corresponding to all locations that have the same distance to BS in the y-axis.
Hence, the error bar shows the maximal and minimal throughput at these locations.
In this paper, the proposed solutions for data-driven MIMO are evaluated under the feedback architecture of 5G but with fixed transmission parameters.
Therefore, the performance loss does not take the saved overheads of pilots and feedback into consideration.

It can be seen from Fig. \ref{single stream} that the performance of both transmission schemes lowers as the distance to BS increases.
However, the gap between the statistic-based solution and CLSM is negligible, since they are both numerically small.
It should be noted that feedback delay significantly affects the performance of CLSM for over 30\% average throughput decrease with only 1 transmission time interval (TTI) delay, which can be seen from Fig \ref{multiple streams}.
Though outperforming CLSM with feedback delay, statistic-based solutions suffer large performance degradation when multiple data streams are transmitted, with up to 22.33\% and 25.19\% throughput loss, respectively.
Thus, it is of greater value to study the multi-stream transmission of data-driven MIMO, which is the focus of this paper.

\begin{figure}[t]
    \begin{center}
        \includegraphics[width=0.45\textwidth]{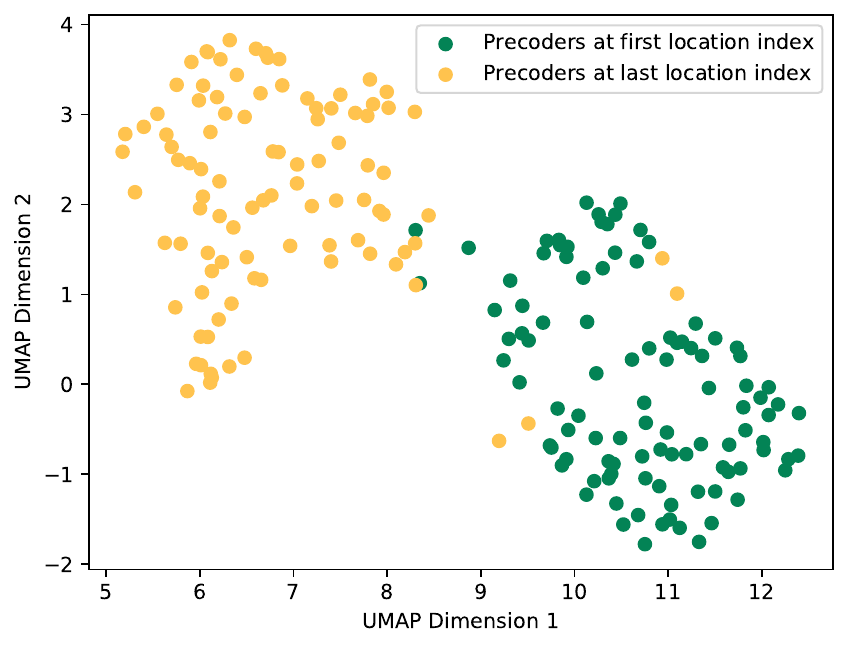}
        \caption{The visualization of latent variables of VAE based on UMAP corresponding to the optimal precoders at the first and last location indexes in $\mathcal{Q}^{3,100}_{\text{fixed}}$.}
        \label{UMAP}
    \end{center}
\end{figure} \par

\subsubsection{Verification of Tricks on VAE}

We test the performance of proposed tricks for VAE-based solutions using 20 samples of locations suitable for transmitting four data streams and evaluate the average performance on all samples of those locations.
Before that, we intend to demonstrate the interpretability of the VAE-based solution.
Specifically, we use the uniform manifold approximation and projection (UMAP) method to visualize the latent variables of VAE \cite{mcinnes2018umap}.
UMAP is widely used for dimension reduction and can project the latent variables into a 2D space, which is shown in Fig. \ref{UMAP}.
Each dot represents the latent representations of the optimal precoder at a certain time index, thus 100 dots of the same color refer to those at one location index.
In the projected 2D space of UMAP, precoders at the same location are mostly entangled with each other while precoders among different locations present diverse representations.
These properties assist to choose the representative latent variables at observed locations and infer them at unobserved locations.

\begin{figure}[htbp]
    \begin{center}
    \subfigure[Non-unitary matrices]
    {
        \label{Non unitary}
        \includegraphics[width=0.45\linewidth]{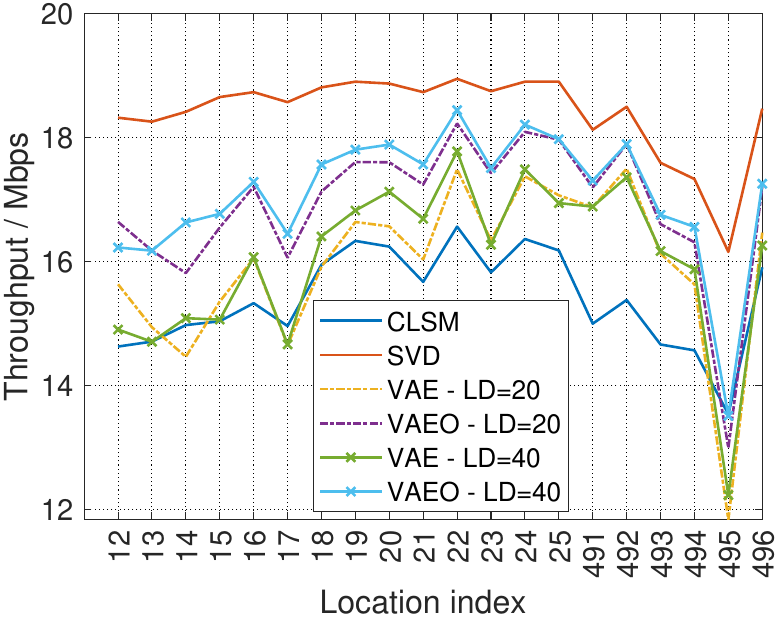}
    }
    \subfigure[Unitary matrices]
    {
        \label{Unitary}
        \includegraphics[width=0.45\linewidth]{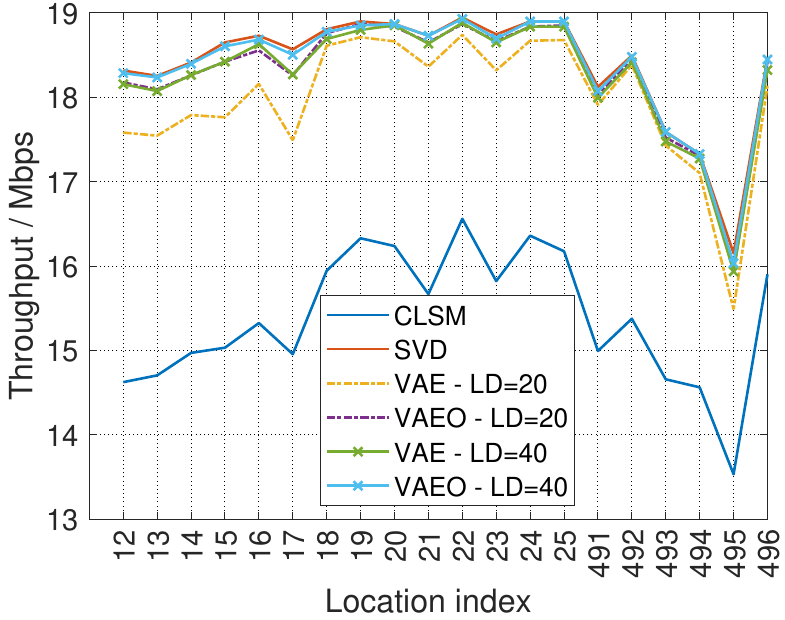}
    }
    \end{center}
    \caption{Average throughput comparison among different transmission schemes in time domain on 20 location samples in $\mathcal{Q}^{4,100}_{\text{fixed}}$ when the precoder dataset consists of non-unitary matrices or unitary matrices.}
    \label{performance on 20 samples}
\end{figure} \par

\begin{figure}[htbp]
    \begin{center}
    \subfigure[SVD]
    {
        \label{SVD}
        \includegraphics[width=0.3\linewidth]{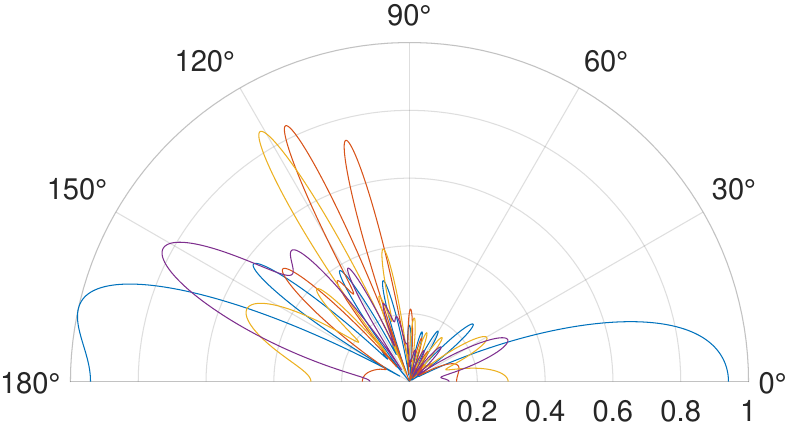}
    }
    \subfigure[Non-unitary matrices as dataset of VAEO]
    {
        \label{orthogonalization-non unitary}
        \includegraphics[width=0.3\linewidth]{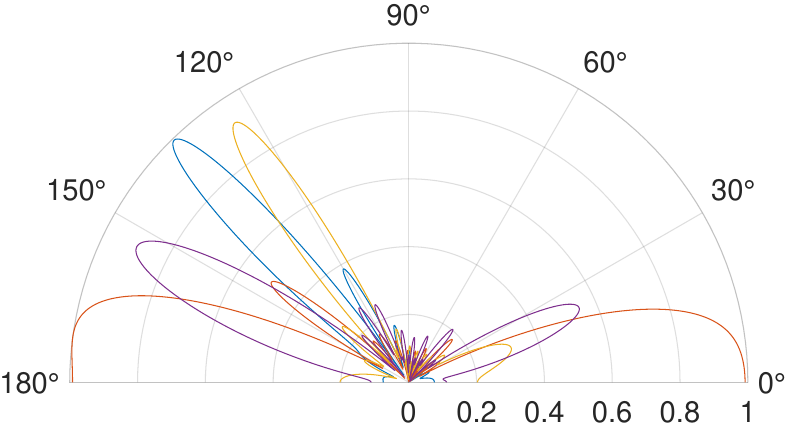}
    }
    \subfigure[Unitary matrices as dataset of VAEO]
    {
        \label{orthogonalization-unitary}
        \includegraphics[width=0.3\linewidth]{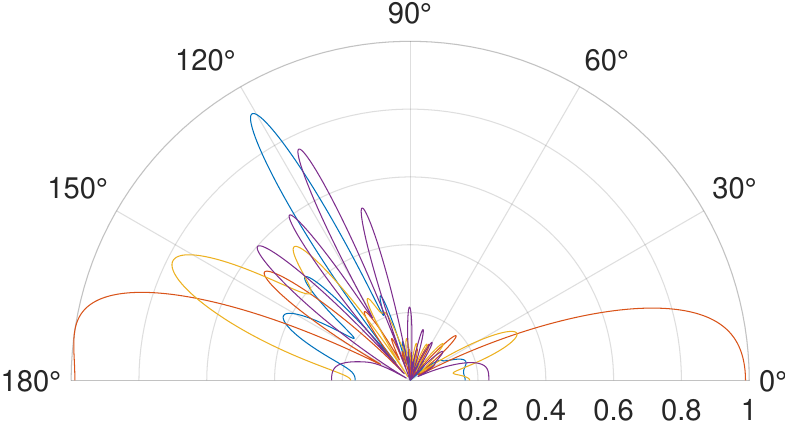}
    }
    \end{center}
    \caption{Radiation patterns comparison among the original SVD precoder at first indexes of time and space domains and its recovered ones from VAEO when different kinds of precoder datasets are employed.}
    \label{radiation pattern}
\end{figure} \par

Fig. \ref{performance on 20 samples} shows the average throughput comparison in time domain when the precoder dataset is made up of non-unitary matrices or unitary matrices.
LD in Fig. \ref{performance on 20 samples} refers to the latent dimension, and VAEO indicates VAE with orthogonalization operation.
Employing precoders of unitary matrices as the precoder dataset shows significantly better reconstruction performance of the original SVD-based precoders than directly using the energy normalized precoders since the values of normalized ones are too small.
VAE of higher latent dimension has greater reconstruction performance for it can better capture the characteristics of precoders.
Besides, the orthogonalization operation offers large performance gain since it keeps the multiple streams orthogonalized to each other to reduce the interference among them.
The radiation patterns in Fig. \ref{radiation pattern} intuitively illustrate that using precoders consisting of unitary matrices along with orthogonalization can reconstruct the original beams well.
The conclusion can be drawn from the above that VAEO of 40 latent dimensions has the best performance when $\mathfrak{r} = 4$, then we use it as the standard to find the representative precoder of rank 4 according to the methods described in Section \ref{Fixing of Transmission Parameters}.

\begin{figure}[htbp]
    \begin{center}
        \includegraphics[width=0.45\textwidth]{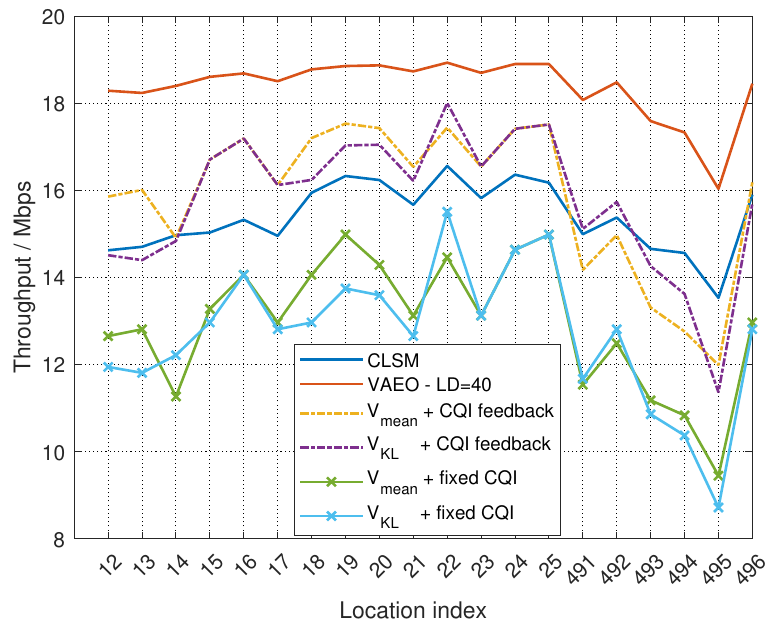}
        \caption{Average throughput comparison among different transmission schemes in time domain on 20 location samples in $\mathcal{Q}^{4,100}_{\text{fixed}}$.}
        \label{Fixed V+CQI on 20 samples}
    \end{center}
\end{figure} \par

Fig. \ref{Fixed V+CQI on 20 samples} demonstrates the performance loss in time domain when we fix the precoder and CQI.
On average, throughput drops 13.48\% and 14.09\% when the precoder is fixed compared with the VAEO of latent dimension size 40, and 29.44\% and 30.77\% when the CQI is then fixed.
The two methods of choosing the representative precoder exhibit similar performance, yet they do not perform well when the channel condition is poor.
Then we focus on all the location samples that employ precoders of rank 4 for all time indexes from the results of SVD.
We find that RIs derived from the statistic-based method are not always four at these locations, thus we divide them into two parts: locations where the fixed RI are both four from statistic-based and VAE-based solutions and the rest of them.

\begin{figure}[htbp]
    \begin{center}
    \subfigure[Same fixed RI obtained from CLSM and SVD]
    {
        \label{Same fixed RI}
        \includegraphics[width=0.45\linewidth]{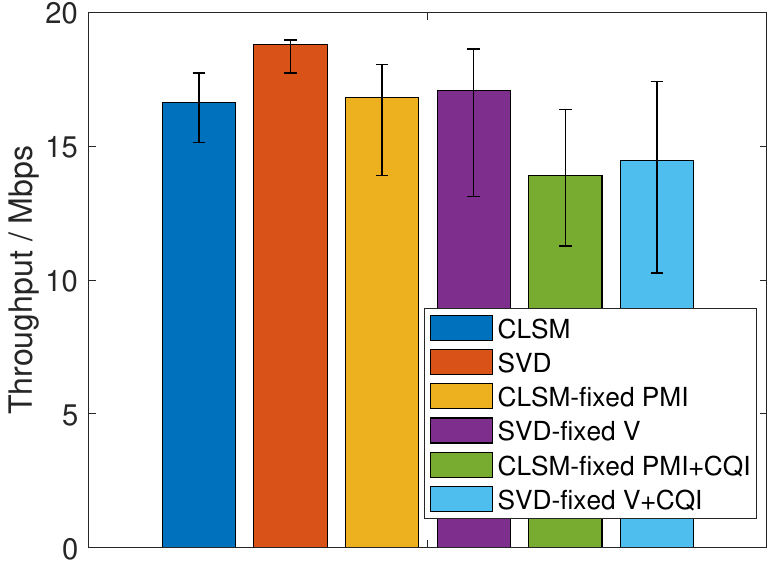}
    }
    \subfigure[Different fixed RIs obtained from CLSM and SVD]
    {
        \label{Different fixed RIs}
        \includegraphics[width=0.45\linewidth]{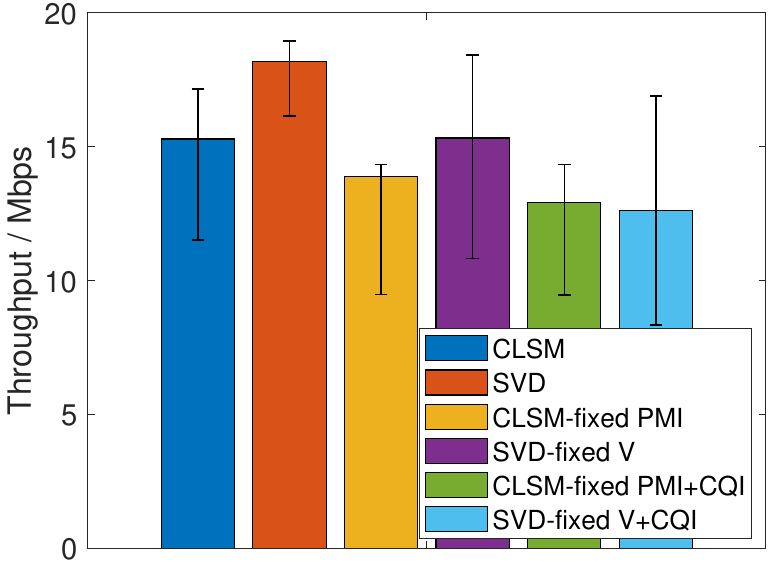}
    }
    \end{center}
    \caption{Average throughput comparison among different transmission schemes on location samples in $\mathcal{Q}^{4,100}_{\text{fixed}}$.}
    \label{4 comp}
\end{figure} \par

The fixed PMI and CQI of the codebook-based approach from CLSM in Fig. \ref{4 comp} is chosen according to statistic-based solution 3.
$\mathbf{V}_{\text{mean}}$ is chosen to be the method to fix the SVD precoder in Fig. \ref{4 comp}.
As demonstrated in Fig. \ref{Same fixed RI}, the performance of the VAE-based solution outperforms the statistic-based solution if their fixed RIs are the same, since the former can better form the beams towards the multipath signals.
CLSM based on the codebook has a conservative preference for the selection of transmission parameters, and RIs of it are 3 and 4 in Fig. \ref{Same fixed RI}, leading to relatively small throughput fluctuations.
Therefore, precoders with a fixed rank of 4 can outperform CLSM in average throughput, though bringing in more fluctuations.
The VAE-based solution is expected to outperform the statistic-based solution if the former has a higher fixed RI since it has more data streams to transmit.
However, the simulation results prove the other way around, as shown in Fig. \ref{Different fixed RIs}.
Although the VAE-based solution outperforms if only precoders are fixed, it has severe performance degradation when CQI is then fixed because it brings higher CQI fluctuation due to higher RI.
It raises the key to this problem: choosing the proper fixed RI at each location.
The conservative choice made by the codebook-based approach can reduce the performance loss brought by fixed transmission parameters when the channel state is not ideal.

\subsubsection{Comparison at All Locations in Training Dataset}

\begin{figure}[htbp]
    \begin{center}
        \includegraphics[width=0.42\textwidth]{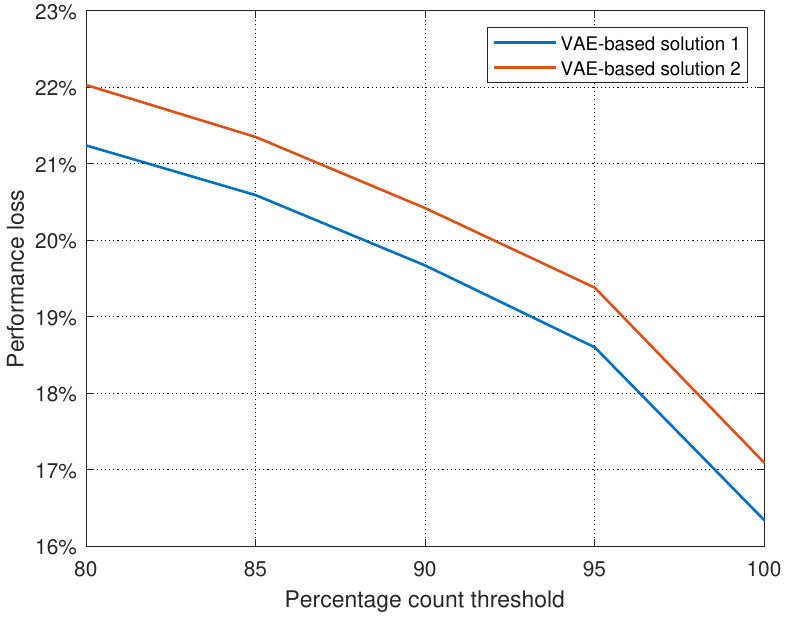}
        \caption{The performance loss of VAE-based solutions compared with CLSM when $\mathfrak{C}_{\text{thold}}$ is changing.}
        \label{threshold}
    \end{center}
\end{figure} \par

\begin{figure}[htbp]
    \begin{center}
        \includegraphics[width=0.45\textwidth]{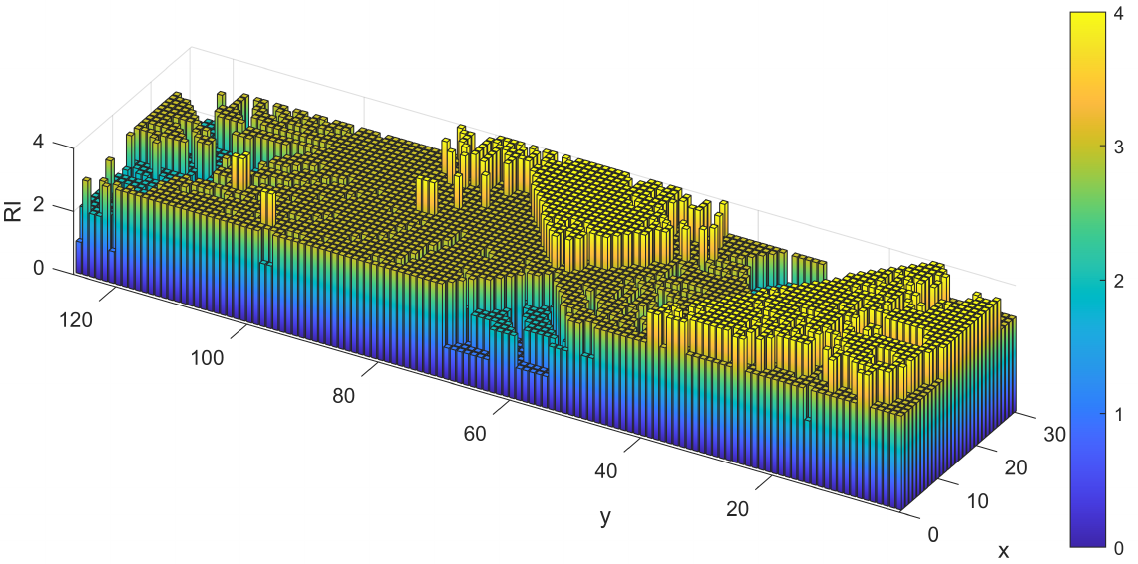}
        \caption{The fixed RIs at locations of UEs in $\mathcal{H}$ when $\mathfrak{C}_{\text{thold}} = 100$.}
        \label{RI bar}
    \end{center}
\end{figure}

As stated before, the proper choice of RI is critical to this problem, and we evaluate the performance loss of VAE-based solutions to CLSM when the criterion for RI selection is changing, which is shown in Fig. \ref{threshold}.
VAE-based solution 1 indicates that the representative precoder of SVD is chosen as $\mathbf{V}_{\text{mean}}$, and solution 2 is $\mathbf{V}_{\text{KL}}$.
It can be seen from Fig. \ref{threshold} that a more conservative choice leads to less performance loss compared with CLSM.
As a result, we set $\mathfrak{C}_{\text{thold}}$ to 100 to choose the fixed RI to obtain the best performance of VAE-based solutions.
Fig. \ref{RI bar} shows the fixed RIs at locations of UEs in the training dataset when we set $\mathfrak{C}_{\text{thold}}$ to 100.
From now on, BS is at coordinate origin to demonstrate the relative locations of UEs to it.
In general, it is more likely to realize higher data stream transmission if UE's location is closer to BS.
However, UEs which are located right under BS are unable to fully utilize the maximal streams to transmit data, since the channel matrices tend to be ill-conditioned.

 \par

\begin{figure}[htbp]
    \begin{center}
        \includegraphics[width=0.45\textwidth]{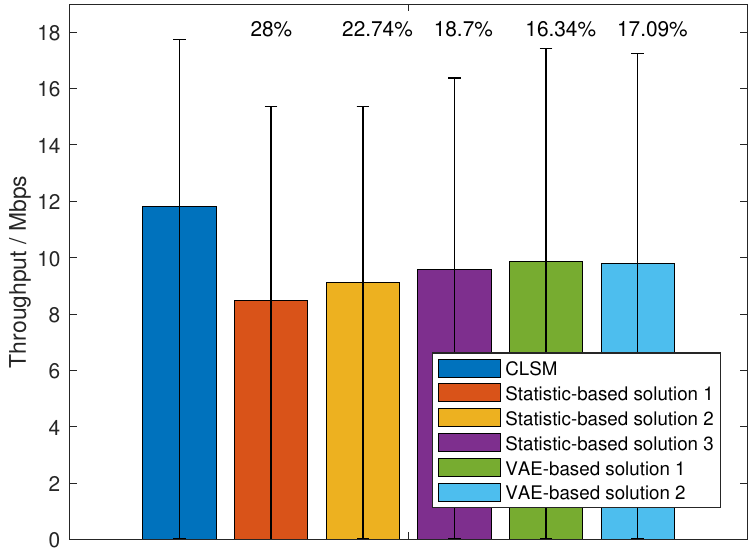}
        \caption{Average throughput comparison among different transmission schemes at locations in $\mathcal{H}$ when $\mathfrak{C}_{\text{thold}} = 100$.}
        \label{trainset_comp}
    \end{center}
\end{figure} \par

Fig. \ref{trainset_comp} shows the average throughput comparison of CLSM and proposed solutions for data-driven MIMO at all locations in $\mathcal{H}$.
Due to the fixed transmission parameters in time domain, the best VAE-based solution exhibits the performance loss of 16.34\%, while 18.7\% for the best statistic-based solution.
VAE-based solutions show superiority after a conservative choice of RI.
Moreover, the performance improvement of VAE-based solutions at locations with the same fixed RI as statistic-based solutions surpasses the reduction at locations with higher RIs.

\begin{figure}[htbp]
    \begin{center}
    \subfigure[Heatmap]
    {
        \label{Heatmap of trainset}
        \includegraphics[width=0.45\linewidth]{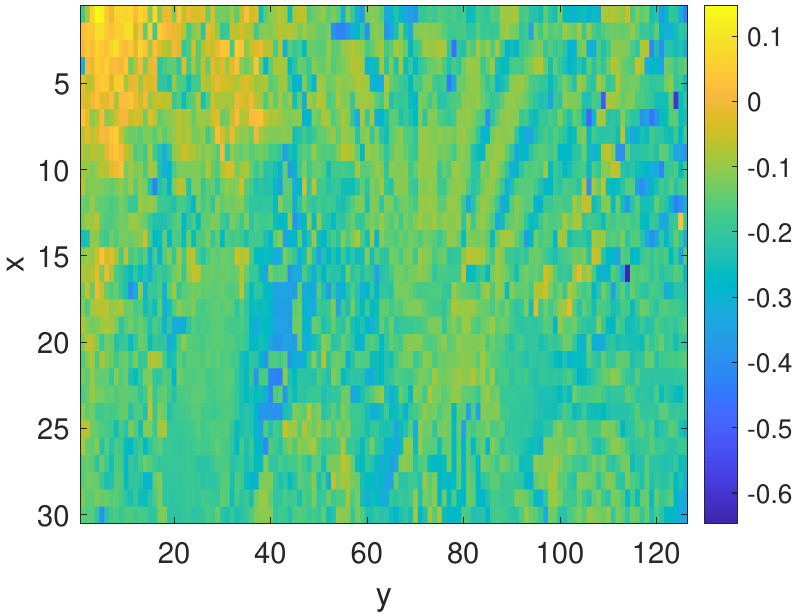}
    }
    \subfigure[Line chart]
    {
        \label{Line chart of trainset}
        \includegraphics[width=0.45\linewidth]{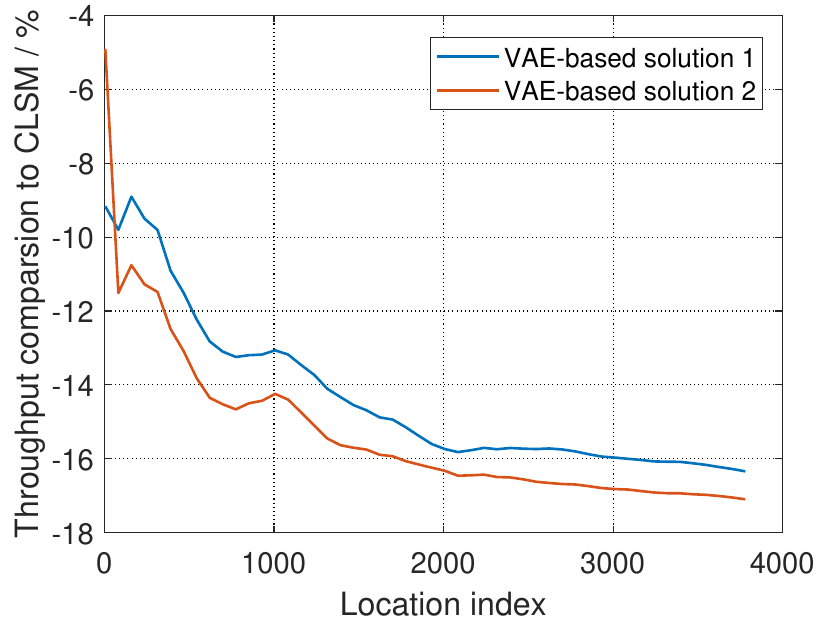}
    }
    \end{center}
    \caption{Average throughput gap ratio of VAE-based solutions compared with CLSM at locations in $\mathcal{H}$.}
    \label{intuitive illustration-trainset}
\end{figure} \par

Fig. \ref{Heatmap of trainset} presents the average throughput gap ratio of VAE-based solution 1 to CLSM, which intuitively shows that VAE-based solution can outperform CLSM if UE is close to BS, although it endures large performance loss at locations where channel states are poor.
It can be seen from Fig. \ref{Line chart of trainset} that the farther the UE is to BS, the worse the performance of VAE-based solutions is compared with CLSM.
The location index starts at the one closest to BS and increases along the x-axis first and then along the y-axis.
In general, VAE-based solution 1 has approximately 1\% better performance than solution 2, except for some users close to BS.

\subsection{Evaluation on Space Domain}

In this subsection, we evaluate the proposed solutions for data-driven MIMO in terms of spatial inference, namely to predict transmission parameters at location samples in the testing dataset $\mathcal{H}'$.

\begin{figure}[htbp]
    \begin{center}
    \subfigure[Statistic-based solution 3]
    {
        \label{Heatmap of CQI-statistic-based solution 3}
        \includegraphics[width=0.45\linewidth]{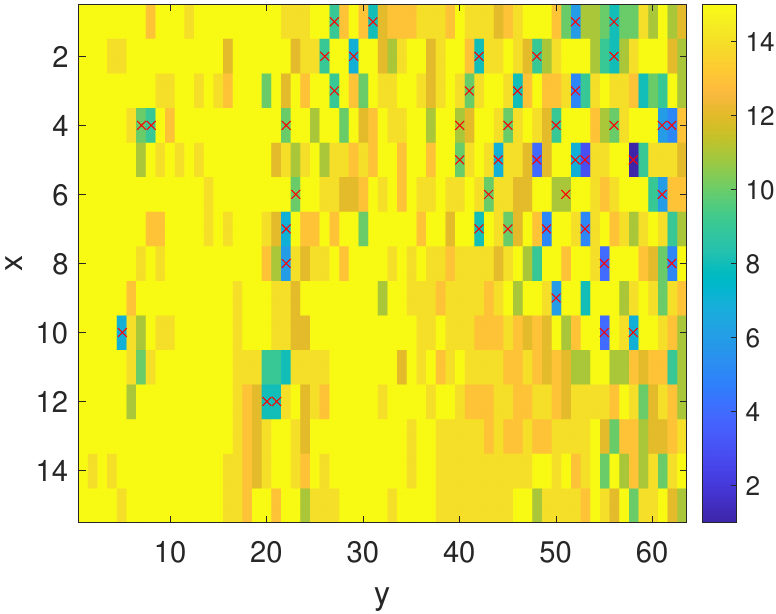}
    }
    \subfigure[VAE-based solution 1]
    {
        \label{Heatmap of CQI-VAE-based solution 1}
        \includegraphics[width=0.45\linewidth]{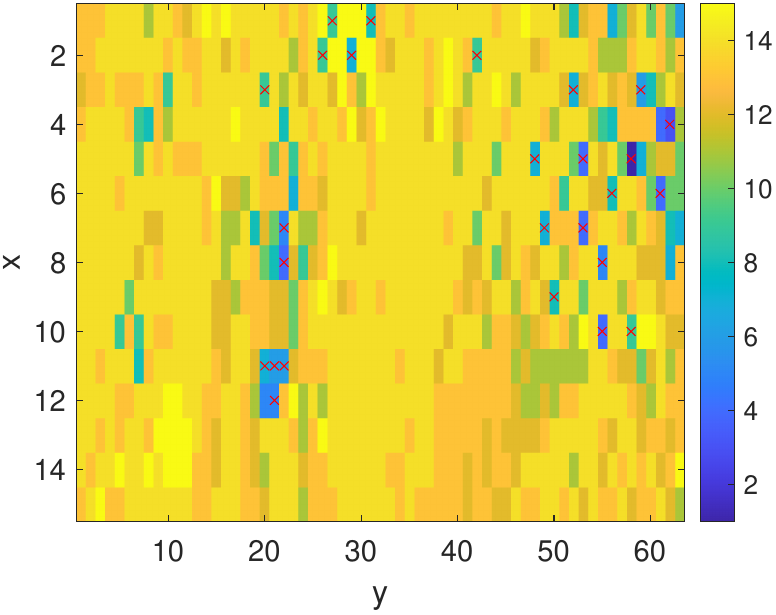}
    }
    \end{center}
    \caption{Heatmap of the round of mean value of true CQIs at locations in $\mathcal{H}'$ when using different methods for data-driven MIMO.}
    \label{CQI heatmap}
\end{figure} \par

Fig. \ref{CQI heatmap} shows the heatmap of the round of mean value of true CQIs if we use the precoders inferred by statistic-based solution 3 and VAE-based solution 1.
We choose the fixed RIs at 4 closest observed locations in $\mathcal{H}$ to determine the interpolated RI at unobserved locations in $\mathcal{H}'$, i.e., $N_{\text{RI}} = 4$.
In general, the fluctuation of CQI is localized and stable, except for some points suddenly experiencing large path loss.
The surge in path loss is difficult to capture by any interpolation method, which leads to zero throughput at certain locations.
Then, locations with average zero throughput are marked with a red cross in Fig. \ref{CQI heatmap}.
The VAE-based solution utilizes GPR for spatial prediction on precoders, which can better catch the underlying relationships among precoders of different locations.
Thus, the VAE-based solution has fewer zero-throughput locations than the statistic-based solution.

\begin{figure}[htbp]
    \begin{center}
    \subfigure[Bar chart]
    {
        \label{testset_comp}
        \includegraphics[width=0.45\linewidth]{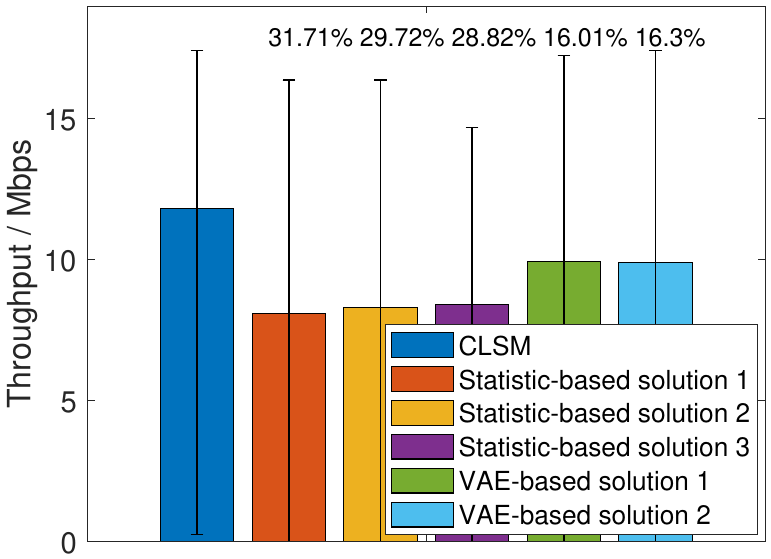}
    }
    \subfigure[Line chart]
    {
        \label{Line chart of testset}
        \includegraphics[width=0.45\linewidth]{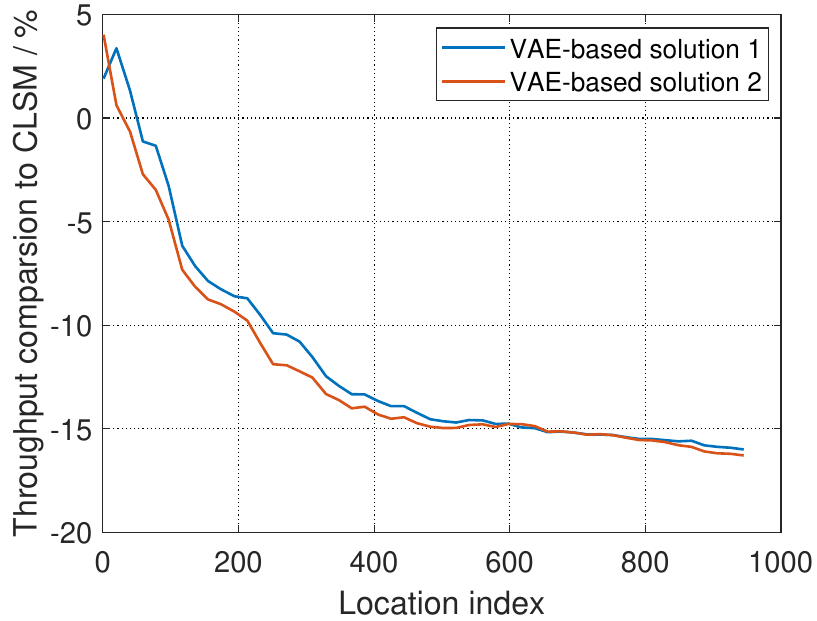}
    }
    \end{center}
    \caption{Average throughput comparison among different transmission schemes at locations in $\mathcal{H}'$.}
    \label{testset}
\end{figure} \par

To sum up, the VAE-based solutions exhibit significant performance gains over statistic-based solutions when predicting the transmission parameters at $\mathcal{Q}'$, as illustrated in Fig. \ref{testset_comp}.
Over 12\% improvement can be made by VAE-based solutions over the best statistic-based solution since VAE combined with GPR can better infer spatial beams at unknown locations.
The performance loss of VAE-based solutions in the testing set in Fig. \ref{testset_comp} is comparable to that in the training set in Fig. \ref{trainset_comp}, which indicates the superiority of VAE-based solutions to statistic-based solutions in spatial inference.
The throughput comparison to CLSM in Fig. \ref{Line chart of testset} shows a similar trend as that in Fig. \ref{Line chart of trainset}.
Overall, the proposed VAE-based solutions only have the performance loss of around 16\% compared with 5G CLSM.
Considering the saved overheads of pilots and feedback, VAE-based solutions for data-driven MIMO can potentially achieve comparable or even higher performance than 5G CLSM.

\section{Conclusion}
\label{Con}
In this paper, we have studied how to implement end-to-end data-driven MIMO for both FD-RAN and future 6G.
It requires no channel feedback and only utilizes geolocation to determine all the transmission parameters based on the mapping learned from historical channel data.
First, an approach based on 5G Type \uppercase\expandafter{\romannumeral1} codebook is proposed, which selects mode values of historical PMIs, RIs, and CQIs and uses nearest-neighbor interpolation for spatial inference.
Then, an SVD-based approach is proposed for performance improvement, where VAE is utilized to choose the representative precoder in time domain.
GPR and NNI are exploited to enable higher spatial prediction accuracy with respect to the precoder and CQI.
Simulation results based on a 5G-compatible link-level simulator and realistic ray-tracing channel data manifest the effectiveness of the proposed approaches, the superiority of the SVD-based approach in spatial inference, and the potential of data-driven MIMO.
In our future work, we will consider the multi-user scenario and utilize other generative artificial intelligence techniques to further realize the capabilities of data-driven MIMO.


\ifCLASSOPTIONcaptionsoff
  \newpage
\fi


\bibliographystyle{IEEEtran}
\bibliography{ref}





\end{document}